\documentclass[useAMS,usenatbib]{mn2e}
\usepackage{times}
\usepackage{url}
\usepackage{graphicx}
\usepackage[usenames]{color}
\usepackage{graphics,graphicx}
\usepackage[usenames]{color}
\DeclareGraphicsExtensions{.pdf,.png,.jpg,.mps,.eps,.ps}
\usepackage{amsmath}
\usepackage{natbib}
\usepackage{pdfpages}
\usepackage{amssymb}
\usepackage{rotating,booktabs}

\def\unit #1{\,{\rm #1}}

\newcommand\kms{\rm \,\unit{km\,s^{-1}}}

\newcommand\cmsqi{\rm \,\unit{cm^{-2}}}
\newcommand\cm{\rm \,\unit{cm}}

\newcommand\cmcubei{\rm \,\unit{cm^{-3}}}

\newcommand\kev{\rm \,\unit{keV}}

\newcommand\lunit{\rm \,erg \,s^{-1}}

\newcommand\xiunit{\rm \,erg\,cm\,s^{-1}}

\newcommand\ledd{L_{\rm \, Edd}}

\newcommand\lbol{L_{\rm \, bol}}
\newcommand\lion{L_{\rm ion}}
\newcommand\labs{L_{\rm abs}}
\newcommand\labstot{L_{\rm abs,Tot}}
\newcommand\labsuta{L_{\rm abs,UTA}}
\newcommand\lxray{L_{\rm Xray}}

\newcommand\msol{M_{\odot}}
\newcommand\medd{\dot{M}_{\rm Edd}}
\newcommand\macc{\dot{M}_{\rm acc}}
\newcommand\mout{\dot{M}_{\rm out}}
\newcommand\moutdust{\dot{M}_{\rm out,dust}}
\newcommand\moutmax{\dot{M}_{\rm out,max}}
\newcommand\moutmin{\dot{M}_{\rm out,min}}
\newcommand\moutmintot{\dot{M}_{\rm out,min,Tot}}

\newcommand\poutmax{\dot{P}_{\rm out,max}}

\newcommand\pouttot{\dot{P}_{\rm out,Tot}}
\newcommand\poutdust{\dot{P}_{\rm out,dust}}
\newcommand\poutdusttot{\dot{P}_{\rm out,dust,Tot}}
\newcommand\poutmin{\dot{P}_{\rm out,min}}
\newcommand\poutmintot{\dot{P}_{\rm out,min,Tot}}
\newcommand\pout{\dot{P}_{\rm out}}
\newcommand\pabs{\dot{P}_{\rm abs}}
\newcommand\prad{\dot{P}_{\rm rad}}
\newcommand\pabstot{\dot{P}_{\rm abs,Tot}}

\newcommand\Ekmax{\dot{E}_{\rm K,max}}
\newcommand\Ekdust{\dot{E}_{\rm K,dust}}
\newcommand\Ekdusttot{\dot{E}_{\rm K,dust,Tot}}

\newcommand\Ektot{\dot{E}_{\rm K, Tot}}
\newcommand\Ekmin{\dot{E}_{\rm K,min}}
\newcommand\Ekmintot{\dot{E}_{\rm K,min,Tot}}
\newcommand\Ek{\dot{E}_{\rm K}}
\newcommand\Ekout{\dot{E}_{\rm K,out}}

\newcommand\mbh{M_{\rm BH}}

\newcommand\nh{ N_{\rm H}}

\newcommand\pc{\unit{pc}}

\newcommand\kpc{\unit{kpc}}

\newcommand\mpc{\unit{Mpc}}

\newcommand\ev{\unit{\, eV}}

\newcommand\rmin{r_{\rm min}}
\newcommand\rmax{r_{\rm max}}
\newcommand\rs{r_{\rm s}}
\newcommand\rtorus{r_{\rm torus}}

\newcommand\rdust{r_{\rm dust}}

\newcommand\vout{v_{\rm out}}

\newcommand\vf{V_{\rm f}}

\newcommand\lgxi{\rm \, log\xi}

\newcommand\chandra{{\it Chandra} {}}

\newcommand\xmm{{\it XMM-Newton} {}}

\usepackage{ulem}
\usepackage{graphicx}

\title[Warm absorbers in Xrays II (WAX-II)]{Warm Absorbers in X-rays (WAX), a comprehensive high resolution grating spectral study of a sample of Seyfert Galaxies: \\ II. Warm Absorber dynamics and feedback to galaxies.}

\author[Laha et al.]{Sibasish Laha$^{1}$\thanks{sib.laha@gmail.com, s.laha@qub.ac.uk}, Matteo Guainazzi$^{2,3}$\thanks{Matteo.Guainazzi@sciops.esa.int}\thanks{mguainaz@astro.isas.jaxa.jp}, Susmita Chakravorty$^{4}$, Gulab \ C.\ Dewangan$^{5}$, \newauthor and, Ajit \ K.\ Kembhavi$^{5}$. \\
$^{1}$School of Mathematics and Physics, The Queen's University of Belfast, Belfast, BT7 1NN, United Kingdom. \\
$^{2}$European Space Astronomy Centre of ESA, PO Box 78, Villanueva de la Canada, 28691, Madrid, Spain.\\
$^{3}$Institute of Space and Astronautical Science, 3-1-1 Yoshinodai, Chuo-ku, Sagamihara, Kanagawa, Japan.\\
$^{4}$Laboratoire d'Astrophysique, Universite Joseph Fourier, CNRS UMR 5571, Grenoble, France. \\
$^{5}$Inter University Centre for Astronomy and Astrophysics, Post bag 4, Ganeshkhind, Pune, India.}

\date{\today}
\begin{document}




\maketitle
\label{firstpage}

\begin{abstract}
 
{This paper is a sequel to the extensive study of warm absorber (WA) in X-rays carried out using high resolution grating spectral data from \xmm{} satellite (WAX-I). Here we discuss the global dynamical properties as well as the energetics of the WA components detected in the WAX sample. The slope of WA density profile ($n\propto r^{-\alpha}$) estimated from the linear regression slope of ionization parameter $\xi$ and column density $\nh$ in the WAX sample is $\alpha=1.236\pm 0.034$.  We find that the WA clouds possibly originate as a result of photo-ionised evaporation from the inner edge of the torus (torus wind). They can also originate in the cooling front of the shock generated by faster accretion disk outflows, the ultra-fast outflows (UFO), impinging onto the interstellar medium or the torus. The acceleration mechanism for the WA is complex and neither radiatively driven wind nor MHD driven wind scenario alone can describe the outflow acceleration. However, we find that radiative forces play a significant role in accelerating the WA through the soft X-ray absorption lines, and also with dust opacity.  Given the large uncertainties in the distance and volume filling factor estimates of the WA, we conclude that the kinetic luminosity $\Ek$ of WA may sometimes be large enough to yield significant feedback to the host galaxy. We find that the lowest ionisation states carry the maximum mass outflow, and the sources with higher Fe M UTA absorption ($15-17\rm \AA$) have more mass outflow rates.}


\end{abstract}


\section{Introduction}

Active galactic nuclei (AGN) are sources of huge energy contributing to panchromatic spectra ranging from radio wavelengths to hard X-rays and Gamma rays. The most commonly accepted picture of an AGN is a super massive blackhole (SMBH) at the centre, accreting matter in the form of a disk. Apart from emitting huge luminosity, AGN also show spectroscopic signatures including absorption features in the UV and X-ray band suggesting the presence of mass outflows in the form of ionised clouds. These mass outflows have been in the eye of active research for the last one and a half decade after the advent of high resolution grating spectrometeres aboard \chandra{} and \xmm{} observatories. These mass outflows may provide us with clues about how the SMBHs interact with the host galaxy and how they co-evolve on cosmic time scale. The main question is whether the mass outflows provide enough material feedback to the host galaxy to solve some of the cosmological puzzles, e.g., chemical enrichment of the host galaxy, the relation between the mass of SMBH and the stellar velocity dispersion, the formation of large scale structures in the early universe, the discrepancy between visible to baryonic mass ratio in the Universe \citep{2001MNRAS.326.1228B}. { The underlying physics of structure formation is that the baryons need to cool before they can gravitationally collapse, otherwise the thermal/radiative forces can balance the inward gravitational pull and stop the condensation. Even when one includes radiative cooling in the cosmological models, the observed fraction of visible to baryonic mass is not obtained. This points to some external heating (shock/outflow/external-radiation) acting on the baryonic matter which prevents it from cooling. The outflows from AGN could be this source of external heating.}

The ionised absorbers, popularly known as the warm absorbers (WA) have been studied extensively, both for individual bright sources like NGC~5548, Mrk~509, NGC~3783, Mrk~766, NGC~4051, MCG-6-30-15, Mrk~704, NGC~5548 etc. \citep{2012A&A...539A.117K,2003ApJ...594..116C,2008RMxAC..32..123K,2006MNRAS.368..903P,2004MNRAS.353..319T,2011ApJ...734...75L,2002A&A...386..427K} as well as in samples of AGN \citep{2015arXiv150600614G,2014MNRAS.441.2613L,2013MNRAS.430.1102T, 2012ApJ...745..107W,2010A&A...521A..57T,2007MNRAS.379.1359M,2005A&A...431..111B}. However, there has not been a conclusive agreement on the origin of these outflows, what drives these outflows and how the feedback from these outflows interact with the host-galaxy.

The first compilation of the existing WA studies using grating data was carried out by \citet{2005A&A...431..111B} where the authors found that the kinetic luminosities of the WA are less than $1\%$ of bolometric luminosity of the source. The authors concluded that the WA may not provide sufficient feedback to the host galaxy to have any impact on its evolution. However, the authors noted that the volume filling factor could play a crucial role in calculating outflow estimates. The authors also found that the mass outflow rates of the WA exceed the mass accretion rates of the SMBH by a factor of several decades. They attributed the origin of the WA clouds to `torus wind', where the torus is a neutral obscuring ring like structure around the SMBH \citep{1995PASP..107..803U}. { \citet{2007MNRAS.379.1359M} studied the WA characteristics in a sample of Seyfert galaxies using high resolution \chandra{} data. The authors found that the mass outflow rates and kinetic energy estimates are crucially dependent on the unknown volume filling factor of the warm absorber clouds. The mass outflow rates of the WA calculated by them could equal the mass accretion rate of the SMBH even for a volume filling factor of just $1\%$. Hence, the warm absorbers could be regarded as a potential feedback source.} A combined UV absorber and X-ray WA study for a sample of six bright Seyfert galaxies were carried out by \citet{2012ApJ...753...75C}. The authors found that even for moderately luminous sources, with bolometric luminosity $\lbol\sim 10^{43-45}\lunit$, the ionised outflows (WA and UV absorbers combined) can provide sufficient feedback to the host galaxies. The authors have obtained well constrained measurements on the distances of the absorbers using the absorption from the excited states (in UV) and absorption variability (in X-rays). Three out of six sources in their sample exhibited WA with kinetic luminosities in the range $0.5-5\%$ of $\lbol$. At least five of the six Seyfert 1s also have mass outflow rates that are $10-1000$ times the mass accretion rates needed to generate their observed luminosities. These results point to the fact that the AGN WA have the potential to deliver significant feedback to the host galaxy.

More recent studies carried out by \citet{2013MNRAS.430.1102T} and \citet{2015arXiv150600614G} involving the ultra-fast-outflows (UFOs), which are high ionisation and high velocity outflows in AGN, found that the UFOs are more important in terms of feedback compared to the warm absorbers. The UFOs yield a kinetic luminosity which is comparable to the bolometric luminosity of the source. The origin of the UFO are presumed to be from the accretion disk winds. A similar study by \citet{2013ApJ...762..103K} carried out comparing the WA, UFO and jet properties from the published literature found that the jets are the most important feedback mechanism as compared to the WA or the UFO. However, feedback from UFOs may be significant in case of high Eddington-ratio sources. The feedback from jets are highly collimated and extend to distances upto \mpc{} and as such they are not able to distribute the material over a large solid angle within the galaxy. Moreover, jets occur in just $10\%$ of the total AGN in nearby universe, which suggests that a more universal phenomemon should be responsible for AGN-Galaxy feedback.



In this work we calculate the WA dynamics using the results from a previous paper of this series \citet{2014MNRAS.441.2613L} (hereafter L1). We have systematically and homogeneously analysed a sample of Seyfert 1 galaxies and characterised their WA properties. The WAX sample is a subsample of the CAIXA (Catalogue of AGN in XMM Archive) sample  which consists all radio quiet X-ray unobscured AGN which were observed by XMM-Newton { by 2008}. This list was cross-correlated with the RXTE X-ray sky survey \citep{2004A&A...418..927R} whereby only those sources were kept in the final list which had a count rate of at least 1 count $\rm s^{-1}$. The final WAX list has 26 sources. The broadband EPIC-pn as well as high resolution RGS data for each source were analysed to ascertain the broadband continuum as well as the WA properties. Simultaneous Optical Monitor (OM) data were used to constrain the UV SED which is crucial for determining the WA properties { \citep[see for e.g.,][]{2013ApJ...777....2L,2013ApJ...776...99R,2013MNRAS.430.2650L} }.

Out of the 26 X-ray type 1 AGN, 17 ($65\%$) sources have at least one warm absorber component. The WAX sample probes the column density of $\rm N_H=10^{20-23.5}\cmsqi$, ionisation parameter of $\rm log\xi=-1 \,to\,3.2$, and outflow velocity of $0-10^4 \kms$. {The ionisation parameter is defined as $\xi=\lion/n_{\rm e} r^2$ where $\lion$ is the ionising luminosity of the source in the energy band 1-1000 Ryd, $n_{\rm e}$ is the electron density of the cloud and $r$ is the distance of the cloud from the central engine.} A total of 33 WA components were detected and the ionisation parameter and the column density were measured with over $3\sigma $ accuracy in most cases. For a few WA components, the velocity could not be constrained. We found that the WA parameters show no correlation among themselves, with the exception of the ionization parameter versus column density. The shallow slope of the $\lgxi$ versus log$v_{\rm out}$ linear regression $0.12 \pm 0.03$ is inconsistent with the scaling laws predicted by radiation or magneto-hydrodynamic-driven winds. Our results suggest also that WA and Ultra Fast Outflows (UFOs) do not represent extreme manifestation of the same astrophysical system. We test these results further in our present work using the dynamical parameters of the WA.


 In this paper we aim at addressing the following questions:
\begin{itemize}
\item Are the WA and the UFOs same type of outflows, and do they have similar origin?

\item The estimation of WA kinetic energy outflow rate. We use the comparison of the quantity against the AGN bolometric luminosity to estimate the AGN feedback to the host galaxy.

\item Is there any correlation between WA absorbed flux and the ionising luminosity or the X-ray luminosity?

\item What is the outflow/launching mechanism of WA and how do they compare with the UFOs and UV absorbers? 

\item What is the density profile of the WA clouds? 
\end{itemize}

This paper is organised as follows: Section \ref{sec:WA-energetics} describes the calculations and the underlying assumptions considered while deriving the dynamical parameters of WA. Section \ref{sec-corr} describes the methods of correlations employed in this work. Section \ref{sec:results} discusses the results followed by conclusion.


\section{Warm absorber energetics}\label{sec:WA-energetics}

In our earlier work L1 we constrained the WA parameters ($\xi$, $\nh$ and $\vout$) for a sample of Seyfert galaxies. In this section we describe the methods we have used to calculate the dynamical parameters of WA from the measured WA parameters. These estimates involve several assumptions. We discuss the implications and caveats of such assumptions in this section. The errors on the derived parameters are calculated following the methods of statistical error propagation from the errors on the measured parameters.

\subsection{WA radial distance}
\label{subsec:radial-dist}

The radial distances of the WA clouds from the central engine are best estimated using the variability studies of WA in response to the continuum changes \citep[see for e.g.,][and the references therein]{2013ApJ...776...99R,2007ApJ...659.1022K,2004ApJ...611...68K,1999ApJ...512..184N}. Such a study is beyond the scope of this paper, and will be done in a future work. In this paper we have calculated the radial distances using three different considerations. 

Firstly, we consider the launch radius $\rmin$ using the virial relationship which assumes that the observed velocity of the WA is the radial escape velocity at that distance from the central engine. 
 
\begin{equation}
r_{\rm min}=\frac{GM}{v_{\rm out}^2}.\label{equ:rmin}
\end{equation}

\noindent As a caveat we must note that in a few cases of X-ray binaries it has been seen that the ouflowing gases have velocities much less than the escape velocity at that radius \citep{2006Natur.441..953M}. 

\begin{figure*}   \centering
\hbox{
 \includegraphics[width=8.5cm,angle=0]{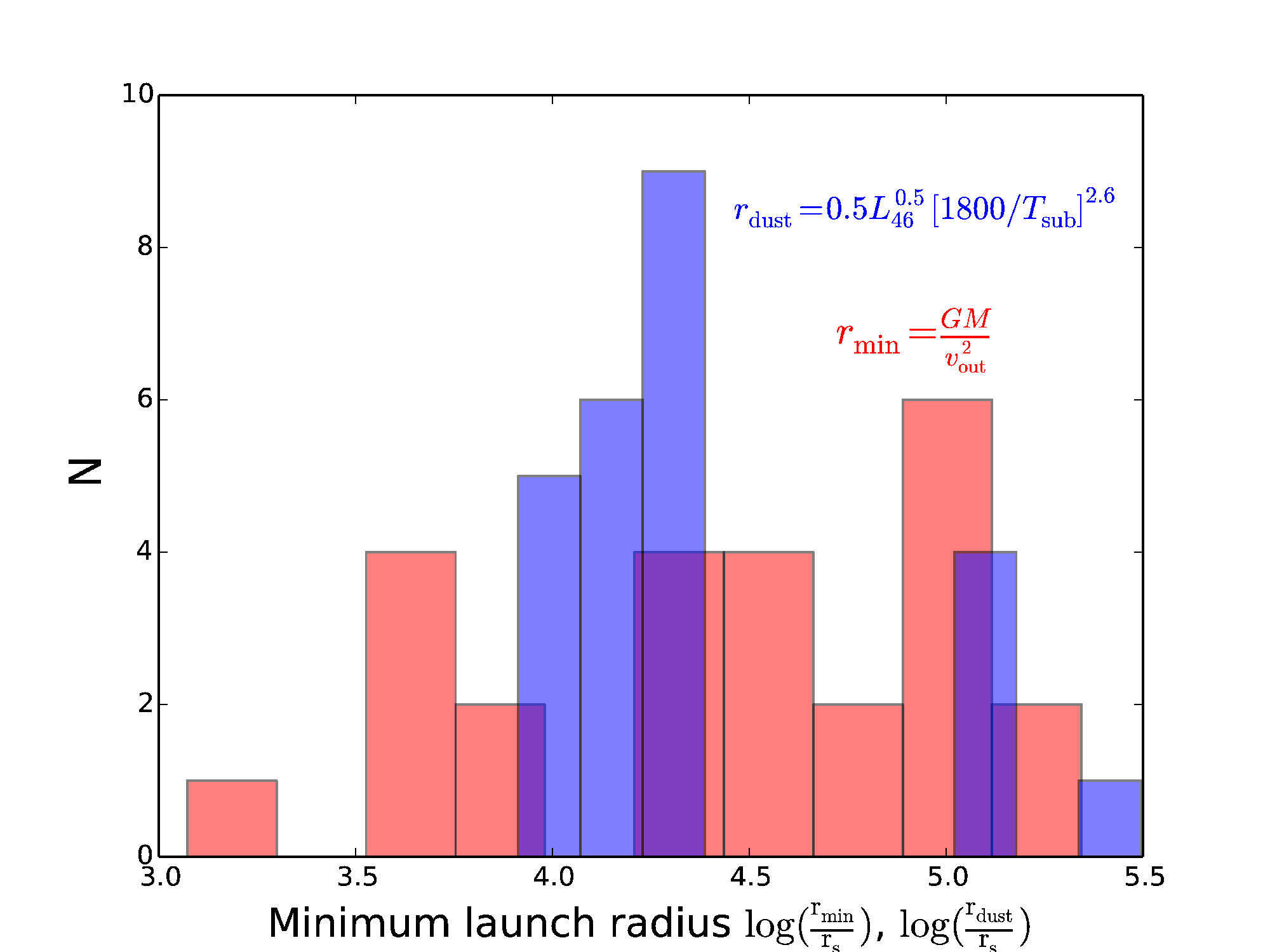}
\includegraphics[width=8.5cm,angle=0]{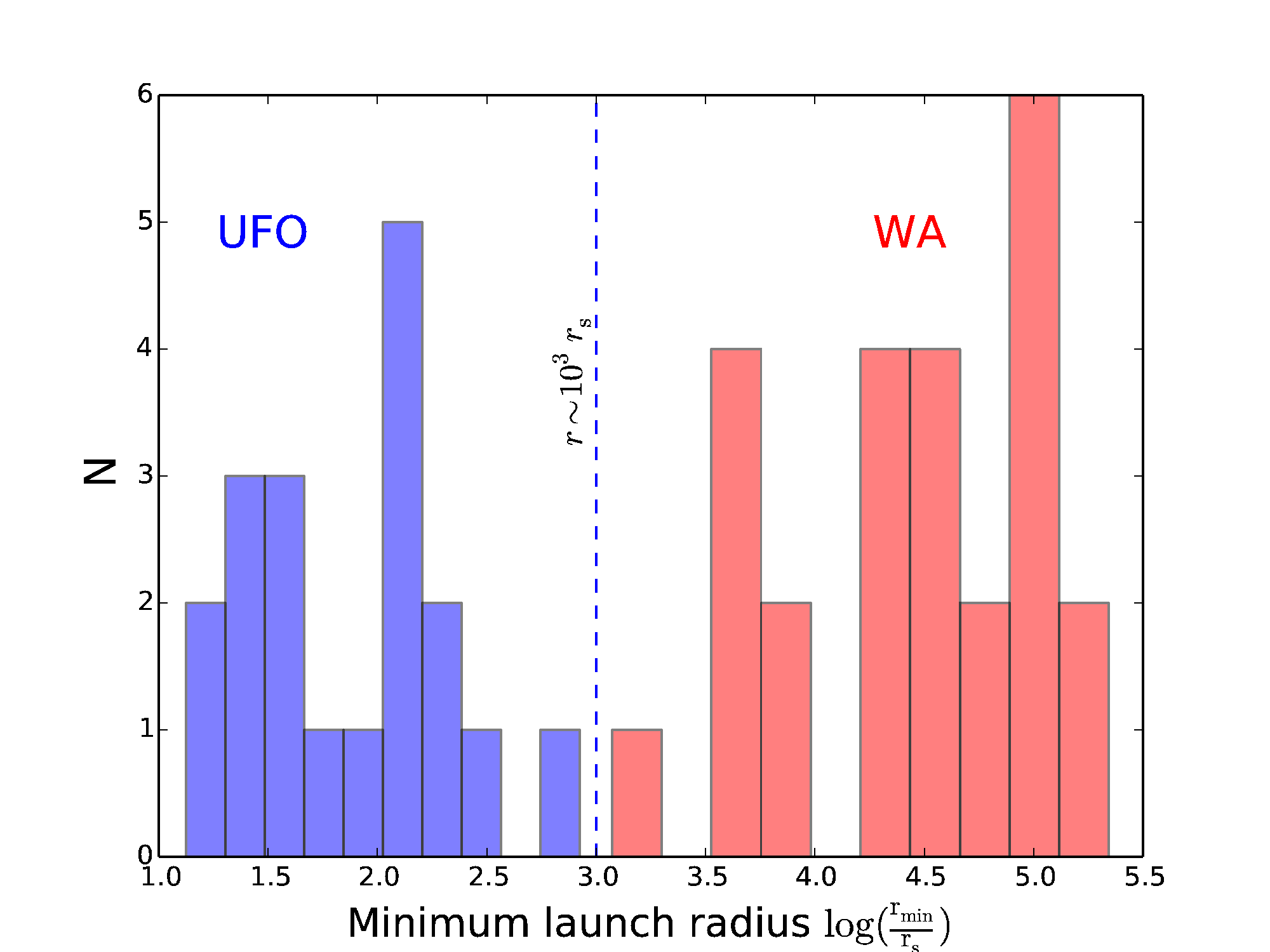}

 }
\caption{{\it LEFT:}The distribution of $\rdust$ (in blue) and $\rmin$ (in red), both scaled with the Schwarzschild radius of SMBH $\rs$. See Section \ref{subsec:radial-dist} for details. {\it RIGHT:} The distribution of $\rmin$ of the UFO (in blue) and WAX (in red), again scaled with respect to $\rs$. We find that the UFO and the WA are separated at a distance $10^3 \rs$.}
\label{fig:rdust}

\end{figure*}

\begin{figure*}   \centering

 \includegraphics[width=8.5cm,angle=0]{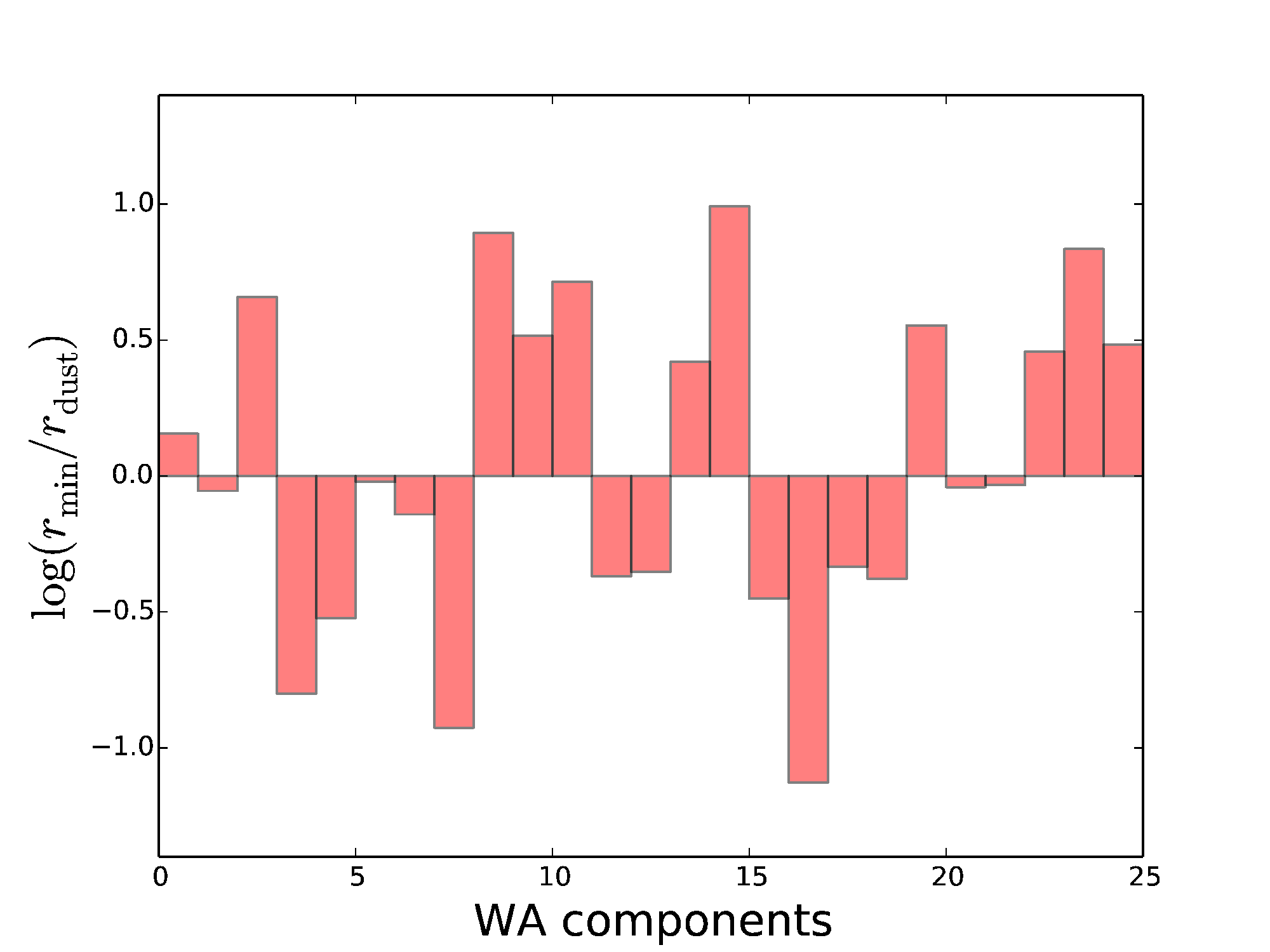}

\caption{The distribution of $\log(\rmin/\rdust)$ for all the 25 individual WA components of the WAX sample. We find that the difference does not exceed an order of magnitude.}
\label{fig:rmin-by-rdust}

\end{figure*}

Secondly, the launch radius can also be calculated by assuming that the WA originates at the base of the broad line region (BLR). A recent study by \citet{2011A&A...525L...8C} pointed out that the broad line region (BLR) in an AGN could originate as clouds uplifted from the outer parts of the accretion disk where the temperature is $\sim 1000$ K. Beyond the dust sublimation radius the radiation pressure on the dusty clouds can sustain the radiatively driven outflows, while below this radius the dust grains sublimate and there is not enough radiative push on the clouds and they fall back on the disk due to gravity. \citet{2015ARA&A..53..365N} in a recent review has discussed how the dust sublimation radius can give us an estimate of the inner margins of the BLR. \citet{1987ApJ...320..537B} had shown that for AGN dust heated by the central luminous source, a dust sublimation radius can be defined as 

\begin{equation}
\rdust=R_{\rm Sub,graphite}\sim 0.5* L^{0.5}_{46}(1800/T_{\rm sub})^{2.6} f(\theta), 
\end{equation}

\noindent where $L_{46}$ is the bolometric luminosity of the AGN in units of $10^{46}\lunit$ and $T_{\rm sub}$ is the dust sublimation temperature (in K), and $f(\theta)$ is an angular dependent term which is 1 for an isotropic source. Hence, we assume $f(\theta)=1$ for simplicity. In a wind driven by radiation pressure on dusty clouds $\rdust$ is therefore an order-of-magnitude estimate of the launch radius. In Figure \ref{fig:rdust}, left panel we show how the $\rdust$ compares with the virial estimates of $\rmin$ in our sample. { Figure \ref{fig:rmin-by-rdust} shows how the two distance estimates $\rmin$ and $\rdust$ compare for each 25 individual WA.}

A further estimate of the warm absorbing cloud location can be obtained using the geometrical definition of ``cloud". We assume that its thickness $\Delta r$, does not exceed its distance from the ionising source \citep{2001ApJ...561..684K,2005A&A...431..111B}. As $\nh=n_{\rm H} V_f \Delta r$, imposing $\Delta r=r$ yields an upper limit on the cloud launching radius

\begin{equation}\label{equ:rmax}
r_{\rm max}=\frac{L_{\rm ion}V_f}{\xi\nh},
\end{equation}

\noindent where $V_f$ is the volume filling factor of the cloud (see Section \ref{subsec:vol} for details), $\lion$ is the ionising luminosity, $\xi$ is the ionisation parameter, $\nh$ is the column density and $n_{\rm H}$ is the hydrogen density of the cloud. This distance estimate follows from a geometrical argument, which may not always hold. $\rmax$ calculated using Equation \ref{equ:rmax} yields estimates as large as a few $\mpc$ (modulo $V_f$). We discuss the interpretation  of these results in Sec. \ref{subsec:rmax}.

We scale all the distances with the Schwarzschild radius of the super massive blackhole (SMBH),

\begin{equation}
r_{\rm s}=\frac{2GM}{c^2}.
\end{equation}

\noindent While we use the subscript ``min" and ``max" for consistency with the literature published so far, we stress that the above estimates are not commensurable. In this paper we will therefore refrain from calculating ``mean distances" by averaging $\rmin$ and $\rmax$, because the statistical properties of the distribution of this average (and the associated statistical uncertainties) are unknown. So $\rmin$, $\rdust$ and $\rmax$ will be treated separately. We will also treat separately the estimates of outflow properties dependent on these distances.


\subsection{Mass outflow rate}

The estimate of the mass outflow rate depends on the geometry of the outflow which involves the knowledge of volume filling factor, covering fraction, density profile and outflow geometry (conical or spherical or other) of the WA clouds which are still largely unknown. We calculate the outflow parameters following \citet{2007ApJ...659.1022K} who considered a biconical geometry for the WA outflow. The mass outflow rate in such a scenario is given by,

\begin{equation}\label{equ:mout}
 \dot{M}_{\rm out \, (max/min)}=\mu \, \pi \, m_p \, \nh \, \vout \, r_{\rm (max/min)} \, f(\delta,\phi),
\end{equation}

\noindent where $m_p$ and $\nh$ denote the mass of proton and the column density of the WA clouds respectively. $f(\delta,\phi)$ is a factor which depends on the angle between the line of sight to the central source and the accretion disc plane, $\delta$, and the angle formed by the wind with the accretion disc, $\phi$. We have assumed a value $f(\delta,\phi)\sim 1.5$ \citep{2007ApJ...659.1022K,2013MNRAS.430.1102T}. The ratio of the proton to electron abundance ($n_{\rm H}/n_{\rm e}$) is expressed as $\mu$ and we have assumed a Solar value of $1/1.4$. { We compared these mass outflow estimates with a few other different methods in Appendix \ref{appendix} and found that they are similar to within an order of magnitude.}

 In our study, the mass outflow rate is scaled with the Eddington mass accretion rate, which is given by:

\begin{equation}
\dot{M}_{\rm Edd}=\frac{L_{\rm Edd}}{\eta c^2},
\end{equation}

\noindent where $\ledd$ is the Eddington luminosity of the source, given by $1.3\times 10^{38}(M/\msol) \lunit$, and $\eta$ is the accretion efficiency assumed to be 0.1.


\subsection{Kinetic Luminosity}

The kinetic luminosity of the WA cloud is the rate of the kinetic energy emitted by the WA cloud and is given by the equation: 

\begin{equation}
\Ek=L_{\rm KE}=\frac{1}{2} \dot{M}_{\rm out} \, \vout^2.
\end{equation}

\noindent This quantity is scaled by $\ledd$ in our study. The bolometric luminosity $\lbol$ is calculated using the UV-Xray SEDs generated for each source in our earlier study L1 in the energy range $1\ev$ to $250\kev$.


\subsection{Momentum outflow rate}

The momentum outflow rate of the WA cloud is given by:

\begin{equation}
\dot{P}_{\rm out}=\dot{M}_{\rm out} \, \vout,
\end{equation}

\noindent and the radiation momentum rate is given by,

\begin{equation}
\dot{P}_{\rm rad}=\frac{\lion}{c},
\end{equation}

\noindent where $\lion$ is the ionising luminosity of the source measured in the range 1-1000 Ryd. The incident ionising photons interact with the warm absorber clouds and get absorbed or scattered whereby they transfer their momentum to the cloud. The amount of absorption depends on the optical depth offered by the clouds to the impinging photons. The momentum tranferred to the cloud due to the absorption of photons is given by

\begin{equation}
\pabstot=\frac{\labstot}{c}.
\end{equation}

\noindent The total absorbed luminosity $\labstot$ for each source is calculated by adding the $\labs$ for all the WA components detected in a given source. Similarly, for the outflow momentum rates we have considered the total values of the quantitites $\pouttot$ and $\pabstot$ while correlating with the source parameters.


\subsection{Volume filling factor}\label{subsec:vol}

The volume filling factor, $V_f$, like the radial distance, is hard to estimate. Warm absorbers have been predominantly found to be clumpy in the form of discrete clouds in multi-phase ionised medium \citep[see for e.g.,][]{2000ApJ...537..134B,2005ApJ...620..165K,2014MNRAS.437.1776M,2014ApJ...793...61H,2014MNRAS.443.1788C,2014MNRAS.441.2613L}. However, there has not been much consensus on the clumping factor and hence the volume filling factor for WA clouds, which varies from source to source. The previous studies on ionised absorptions in AGN indicate a large range in $V_f$, from $10^{-5}$ to 1 \citep{2005A&A...431..111B,2012ApJ...746L..20K}. Following \citet{2005A&A...431..111B} we calculate the volume filling factor of each of the warm absorbing clouds, assuming that the WA are purely radiatively driven. In such a scenario the total outflow momentum rate of the cloud should equal to the momentum due to absorption and scattering of the ionising luminosity by the WA. This proposition may not always be true, but this gives an order of magnitude estimate of $V_f$.  By equating the WA momentum with the absorbed and scattered momentum we obtain,

\begin{equation}\label{equ:Mdot}
\mout \vout \sim \dot{P}_{\rm abs} + \dot{P}_{\rm scatt},
\end{equation}

\noindent where the scattered momentum is given by,

\begin{equation}
\dot{P}_{\rm scatt}=\frac{\lion}{c}(1-e^{-\tau_{\rm T}})
\end{equation}

\noindent The optical depth for Thomson scattering $\tau_{\rm T}$ is given by

\begin{equation}
\tau_T=\sigma_{\rm T} \nh
\end{equation} 

\noindent Using equations \ref{equ:mout} and \ref{equ:Mdot}, the volume filling factor can be written as,

\begin{equation}\label{equ:cv}
V_f\sim \frac{(\labs+\lion(1-e^{-\tau_{\rm T}}))\xi}{1.23 m_p c \lion f(\delta,\phi) v^2}.
\end{equation}

\noindent We list the volume filling factors of each WA component in Table~\ref{Table:Ek}.


\subsection{The UFO data}

For the ultra fast outflows (UFOs) we have used the kinematic data from \citet{2012MNRAS.422L...1T} and \citet{2013MNRAS.430.1102T}. From their study, we have considered the minimum and maximum distance estimates of the UFO and their corresponding outflow parameters separately as described in Section \ref{subsec:radial-dist}.



\section{Correlation analysis}\label{sec-corr}

\subsection{Choice of correlations}
\label{sec:choice}

There are three WA parameters which are independently measured ($\xi, \nh, \vout$) and four source parameters which are independently derived ($\lion$, $\lbol$, $\lxray$, and $\mbh$) in our previous study L1. The correlation and linear regression analysis involving the source parameters are reported in Table \ref{Table:corr-source}. Correlations involving the WA are carried out in three sets, reported in Tables \ref{Table:corr-rmin}, \ref{Table:corr-rdust} and \ref{Table:corr-rmax}, where the outflow parameters are estimated using the virial distance $\rmin$, the dust sublimation distance $\rdust$ and the maximum distance $\rmax$, respectively. We list the source luminosities in Table \ref{Table:lum}. The Tables \ref{Table:dist}-\ref{Table:Ek} list the calculated distances and the outflow parameters in these three different scenarios along with their statistical errors. The warm absorbed luminosities are listed in Table~\ref{Table:dist}. Wherever the correlation involves the source parameters (for e.g., $\prad$, $\lxray$, $\lbol$) we have added the corresponding WA kinematical quantities for a given source if there are multiple WA. We use the suffix `Tot' whenever that is done.

\subsection{Correlation and linear regression methods}
\label{sec:corr-analysis}

For correlation analysis we have used the freely available Python code by \citet{2012Sci...338.1445N} using the BCES technique \citep{1996ApJ...470..706A} to carry out the linear regression analysis between various WA dynamical quantities. In this method the errors in both variables defining a data point are taken into account, as is any intrinsic scatter that may be present in the data, in addition to the scatter produced by the random variables. We also tested the strength of the correlation analysis using the non-parametric Spearman rank correlation method. There are three WA components whose outflow velocity is consistent with zero. We have ignored those components in our analysis as we are mainly interested in the dynamic properties of WA. We also do not include an additional five WA components whose velocity could not be constrained. So our correlation analysis includes 25 WA components.


\section{Results and Discussion}
\label{sec:results}


\subsection{The density profile of warm absorbers}\label{subsec:amd}

The density profile of the WA clouds can be calculated using the absorption measure distribution (AMD) method described by \citet{2007ApJ...663..799H} and \citet{2009ApJ...703.1346B}. The absorption measure distribution is defined as AMD$=d\nh/d\log\xi$, and measures the distribution of hydrogen column density along the line of sight as a function of the ionization parameter. It is calculated using the different ionic states of the same atom (for e.g., Fe) detected in a given source across the X-ray spectra with different ionization parameter and ionic column densities. { We can estimate the AMD from the WAX sample by assuming that the different WA are the manifestation of the different stages of the same absorber \citep[see also][]{2013MNRAS.430.1102T}. Taking the derivative of the linear regression relation, $\log\nh=a\log\xi +b$, where a and b are the best fit linear regression slope and intercepts respectively (see Figure 5 left panel, and Table 10 of L1), we may write the AMD $\propto \xi^a$, where $a=0.31\pm 0.06$ and $b=20.46\pm 0.11$. The slope of the radial density profile ($n\propto r^{-\alpha}$) from \citet{2009ApJ...703.1346B} is given by $\alpha=(1+2a)/(1+a) \pm \Delta a/(1+a)^2$. By substituting the value of $a$ in this equation we obtain $\alpha=1.236\pm 0.034$, which is similar to that obtained by \citet{2009ApJ...703.1346B} from a small but well studied sample of five Seyfert galaxies ($1<\alpha<1.3$). This conclusion validates our original assumption that the global WA properties in the WAX sample could be considered representative of each individual source outflow evolving through different phases of ionic states and column densities. \citet{2012ApJ...758...70G} had suggested that the information we derive from the absorption lines for individual WA are insufficient to comment on the gross properties of the wind. From our earlier proposition therefore this effect can be mitigated to an extent, if we study the WA properties in a sample, as in WAX. We will discuss the implications of the estimated density profile in Sect. \ref{subsec:acc}.}


\subsection{Origin of warm absorber}\label{subsec:origin}

In this Section we aim at constraining the location where the wind could be launched. The launching scale may provide clues on the outflow launching and acceleration mechanism. The virial estimates of the WA distances, $\rmin$, in the WAX sample range from $10^{16}-10^{18}\cm$ ($\sim 0.01-1\pc$). We calculate an approximate distance of the torus following \citet{2001ApJ...561..684K}, which is given by $\rtorus \sim L^{0.5}_{\rm ion, \, 44}\pc$, calculated considering the dust sublimation temperature and the effects of photo-ionisation. We list the values of $\rtorus$ in Table \ref{Table:dist}. We find that the torus distance is on an average an order of magnitude higher ($\sim 1-10 \pc$) than the estimated $\rmin$ for the WA. This may put the WA in the inner torus region. In the last one and half decade of high resolution spectral study of AGN, several authors have calculated the launching radius of the WA at a distance range spanning six orders of magnitude, from accretion disk winds at $\sim 0.001-0.01\pc$ \citep{2000ApJ...545...63E}, to obscuring torus at $\sim 0.1-1\pc$ \citep{2001ApJ...561..684K,2005A&A...431..111B}, to the narow line region $100-1000\pc$ \citep{2002ApJ...575..732K,2004ApJ...606..151O}. Till date there has not been a consensus about the origin of the WA. However, the distances estimated from the WA response to the continuum changes in some sources have yielded very well constrained upper limits. Table \ref{Table:WA-var} shows a compilation of the WA distances for a few bright well studied sources, Mrk~509, NGC~4051, MR2251-178, NGC~3516, and NGC~3783. These estimates put the WA clouds at a distance range of $\sim 0.001-100\pc$ and sometimes wth a lower limit of $r>71\pc$. From Tables \ref{Table:dist} and \ref{Table:WA-var} we find that the minimum distance estimates of WA $\rmin$ mostly agree with the well constrained distance estimates to within an order of magnitude, except for the source NGC~4051, where $\rmin$ is four orders of magnitude larger than the constrained estimates. 

\citet{2001ApJ...561..684K} have argued that the WA could possibly be the clouds formed from evaporation of the inner edge of the obscuring torus due to the highly intense central radiation. Recent studies by \citet{2011A&A...525L...8C} suggested that the BLR clouds arise beyond the dust sublimation radius, where the thrust on the dust grains could drive the outflow. Thus, the dust sublimation radius could represent a lower limit to the outflow launching radius. In our study we find that the dust sublimation radius for the WAX sources range from $\sim 0.1-1\pc$ (see Table \ref{Table:dist}). From Figure \ref{fig:rdust} left panel, we see that the distribution of the minimum launch radius $\log\rmin$ estimated using the virial relation overlaps with the dust sublimation radius $\log\rdust$ of the sources in the WAX sample. Figure \ref{fig:rmin-by-rdust} shows that the virial distance estimates of the individual WA components agree with the dust sublimation radius for every source to within an order of magnitude. {These results point to the fact that the warm absorbers possibly originate in the inner torus region in the form of heated torus winds as proposed by \citet{2001ApJ...561..684K}, with a significant amount of dust opacity which can drive the outflow.} There has been an extensive debate in the last couple of decades as to whether the WA are dusty or not \citep[see for e.g.,][and the references therein]{2001ApJ...554L..13L,2003ApJ...596..114S,2013MNRAS.431.3127V,2015MNRAS.451...93I,2015MNRAS.449..147T}. \citet{2008MNRAS.385L..43F} had suggested that the presence of dust enhances the radiation pressure for Thomson scattering and thus drives mass outflows even for sub-Eddington luminosity sources. For the WA with low ionisation ($\log\xi<1$) we would expect that dust plays an important role in driving the outflows.

A recent study by \citet{2013MNRAS.433.1369P} showed that the WA could arise out of a Compton cooled shocked wind, when a fast, highly ionised wind, launched very near to the SMBH, loses most of its mechanical energy after shocking against the interstellar medium (ISM). { For the source NGC~4051, with a black hole mass of $\log(M_{\rm BH}/\msol)\sim 6$, the authors have calculated the shock radius to be $\sim 10^{17}\cm$.} From Fig. \ref{fig:rdust} right panel, we find that the launching radius of the UFO and the WA are separated at a radius $\sim 10^3 \rs$. From Table \ref{Table:dist} we find that the Schwarzschild radius $\rs$ for the WAX sample of sources ranges between $10^{12-14}\cm$, which implies that the break in the radius comes at around $10^{15-17}\cm$. This is similar to those found by \citet{2013MNRAS.433.1369P}. The authors also predicted that there should be an abrupt drop in the ionisation parameter at the point of transition between the UFO and the WA. From Fig. \ref{fig:WA-UFO} right panel, we find that at a radius of $\sim 10^3 \rs$, there is a drop in the ionisation prameter between the UFO and the WA. For an adiabatically shocked wind, the drop in the velocity and the ionisation parameter should be of similar magnitude. { Hence, for every source where both WA and UFO have been detected, we have compared the highest ionisation parameter of WA with the highest ionisation parameter of UFO and their corresponding outflow velocities. From \citet{2011ApJ...742...44T} and L1, we find that the ratio of the maximum UFO velocity with the highest WA velocity for the sources that exhibit both UFO and WA is $\sim 10-30$ while the drop in the ionisation parameter is larger $\sim 10-10^3 $. These results point to the fact that the WA could possibly be produced by shock by the UFO on the ISM or the torus.}


\subsection{Warm absorber mechanism of acceleration}\label{subsec:acc}

In our previous study L1, we had found that the slope of $\log\xi$ vs $\log\vout$ in the WAX sample is $0.12\pm 0.03$ which is inconsistent with either purely radiatively driven or magneto-hydrodynamically driven (MHD) wind. In this section we investigate the possibility of different driving mechanisms on the basis of the correlations we obtain between the different outflow parameters of WA. 

\begin{figure*} 
  \centering

\vbox{  

\hbox{
 \includegraphics[width=7.5cm,angle=0]{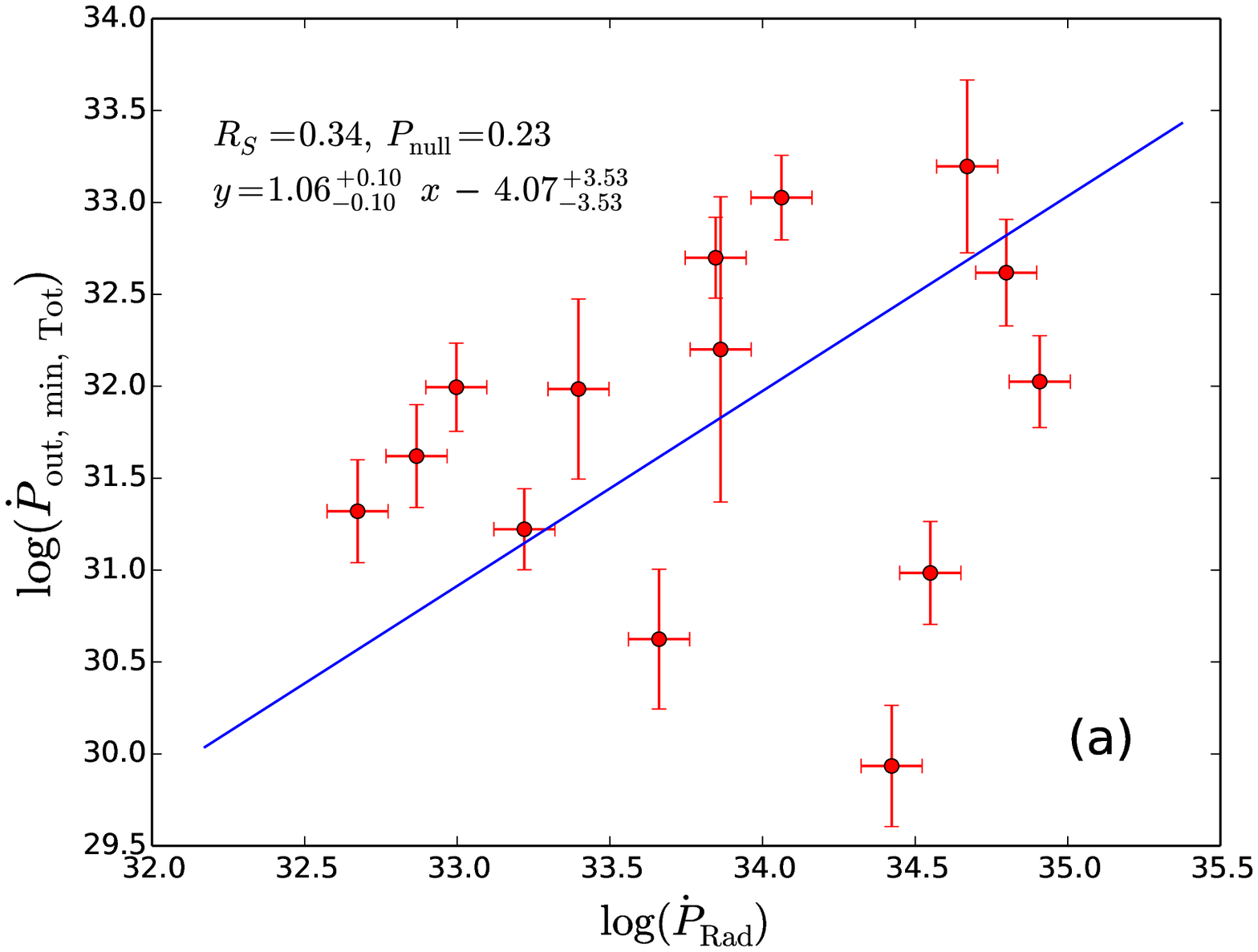}
 \includegraphics[width=7.5cm,angle=0]{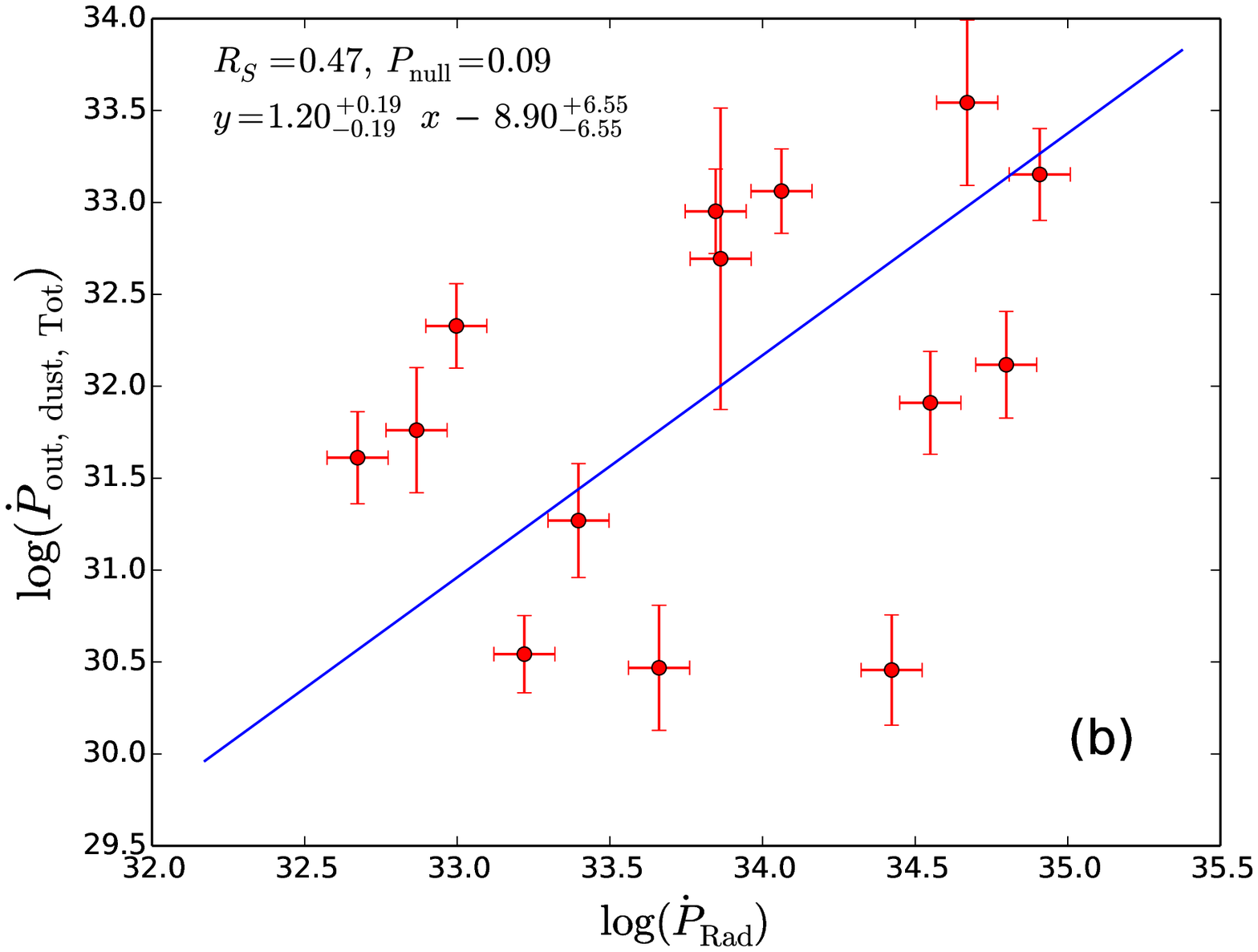}
 }

\hbox{
 \includegraphics[width=7.5cm,angle=0]{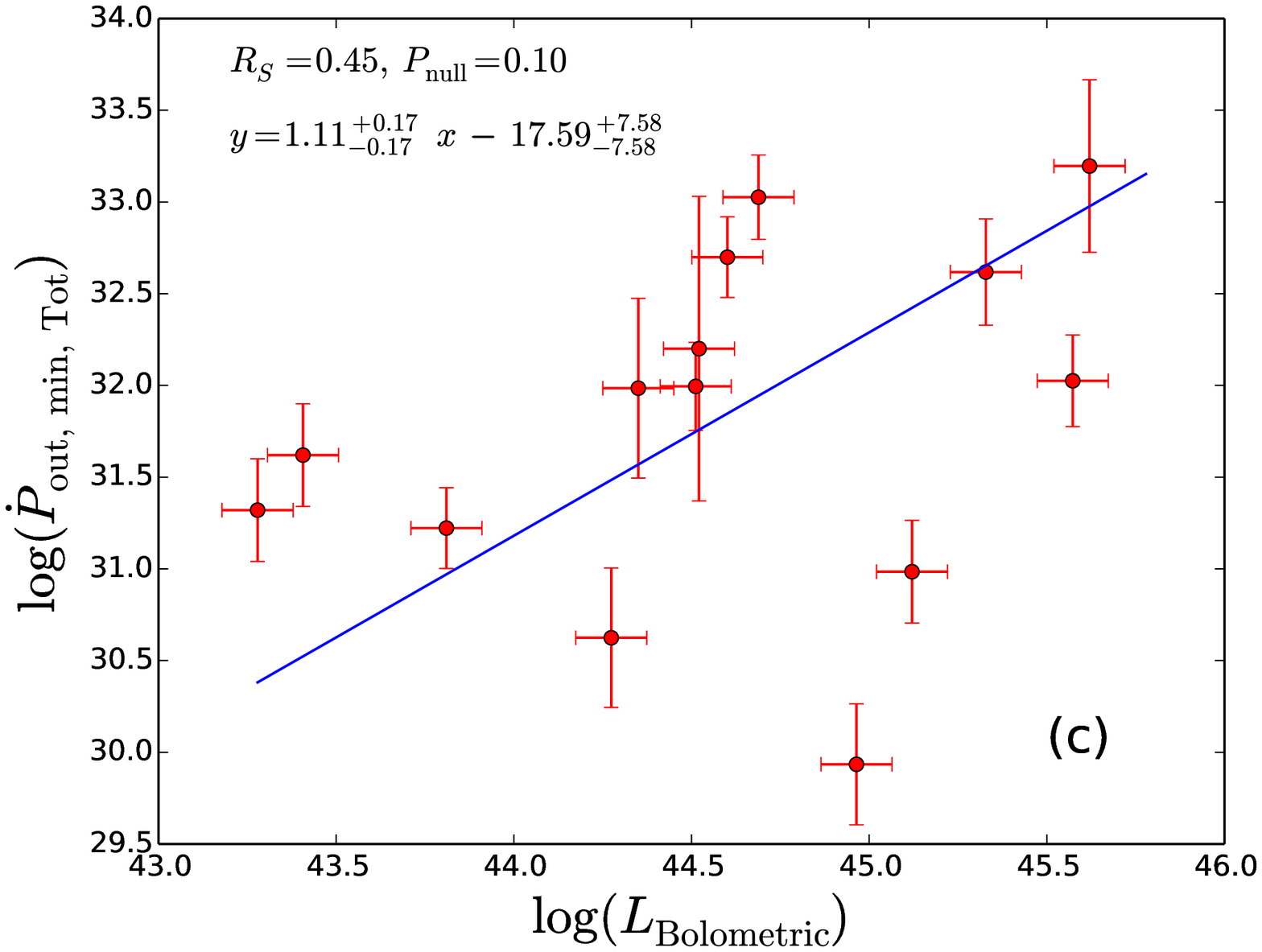}
 \includegraphics[width=7.5cm,angle=0]{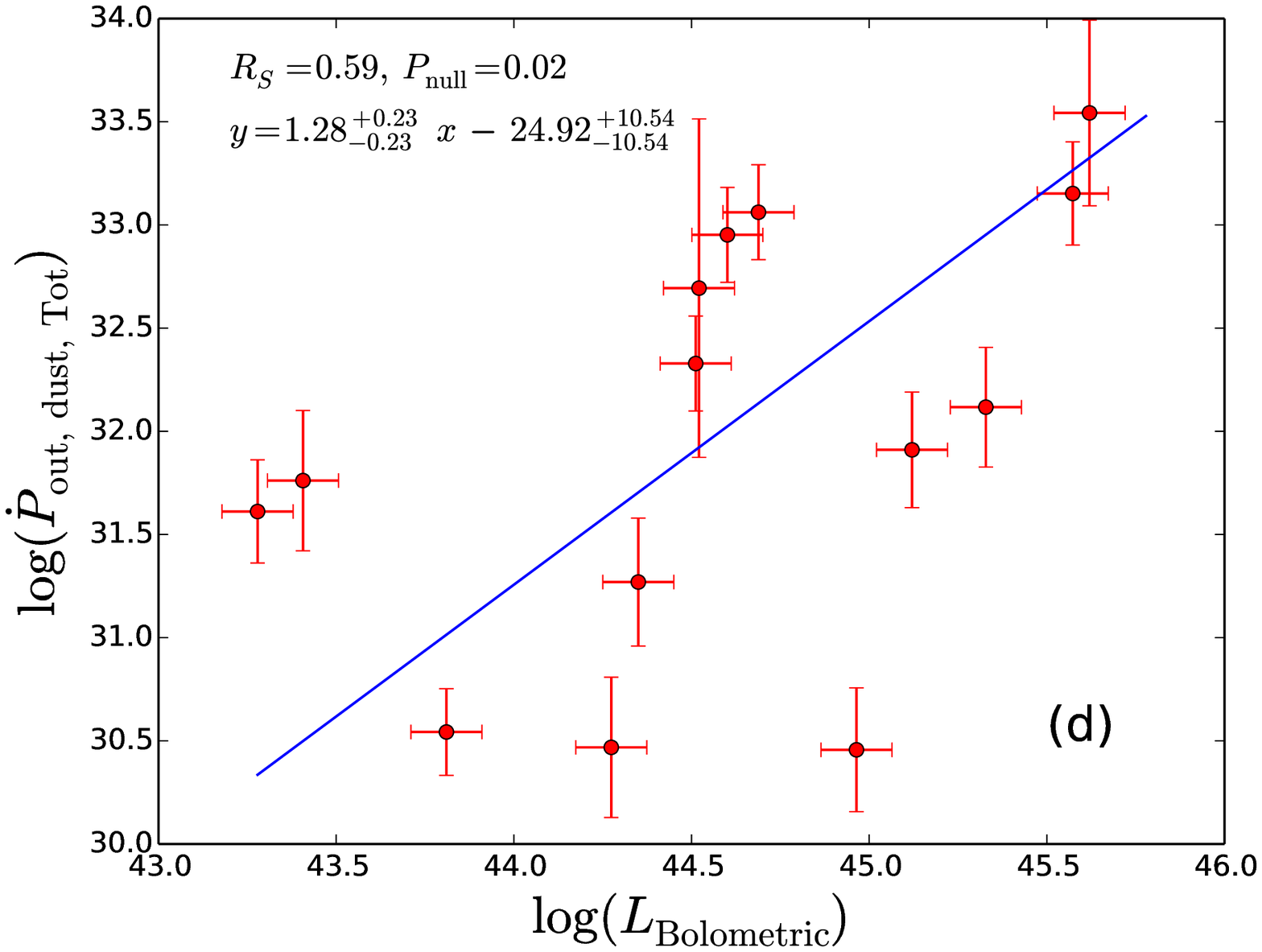}
 }

\hbox{
 \includegraphics[width=7.5cm,angle=0]{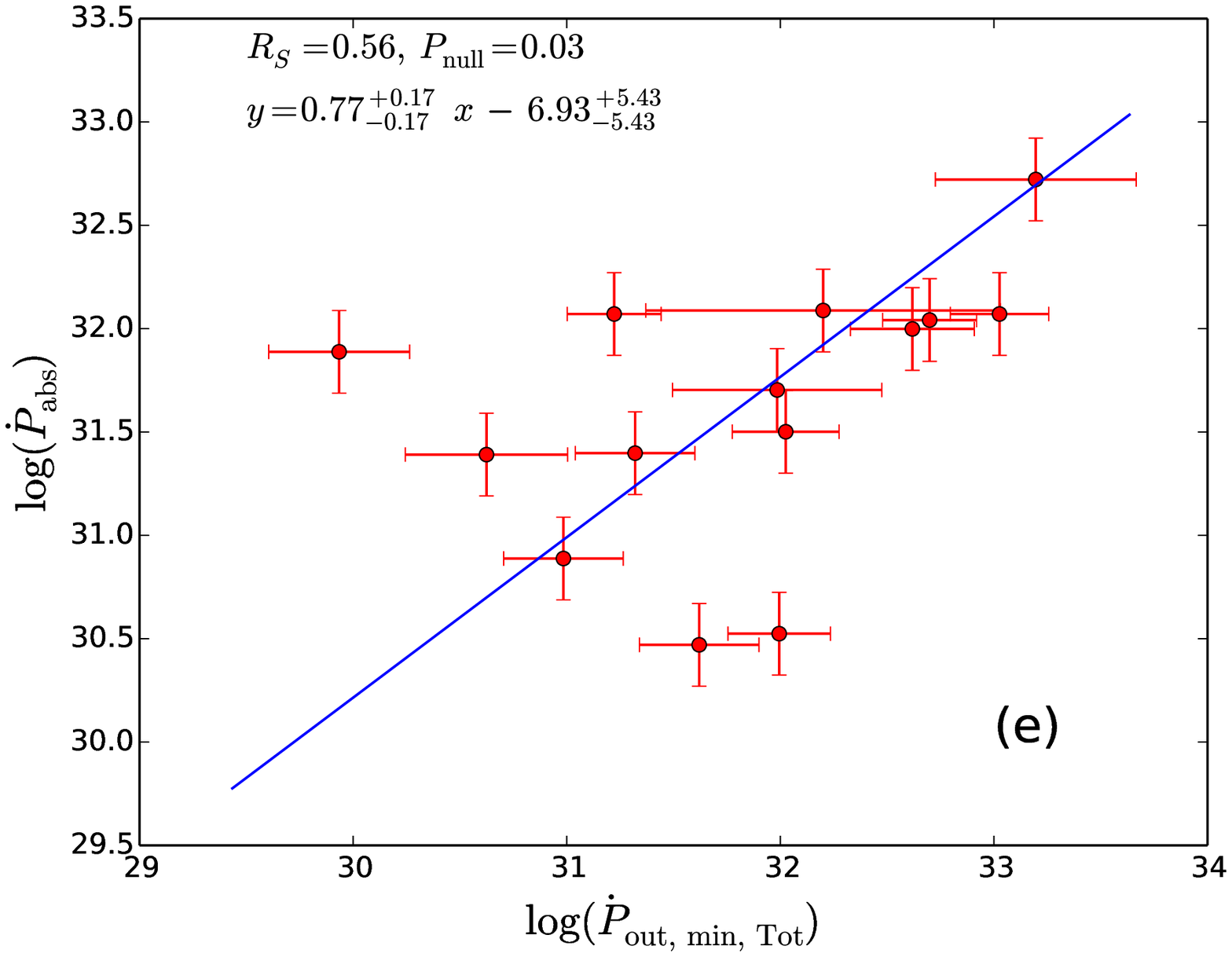}
 \includegraphics[width=7.5cm,angle=0]{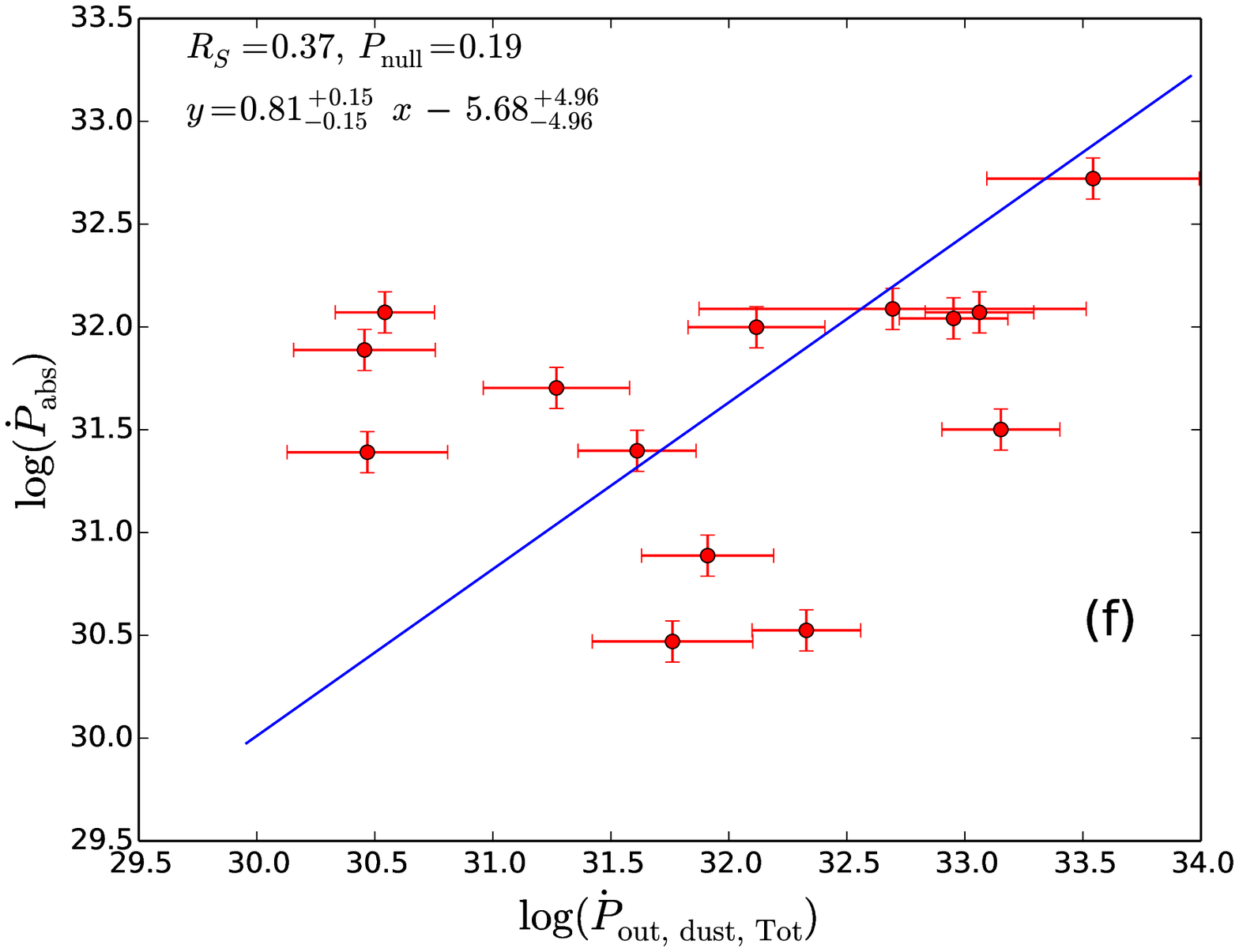}
 }

\hbox{
 \includegraphics[width=7.5cm,angle=0]{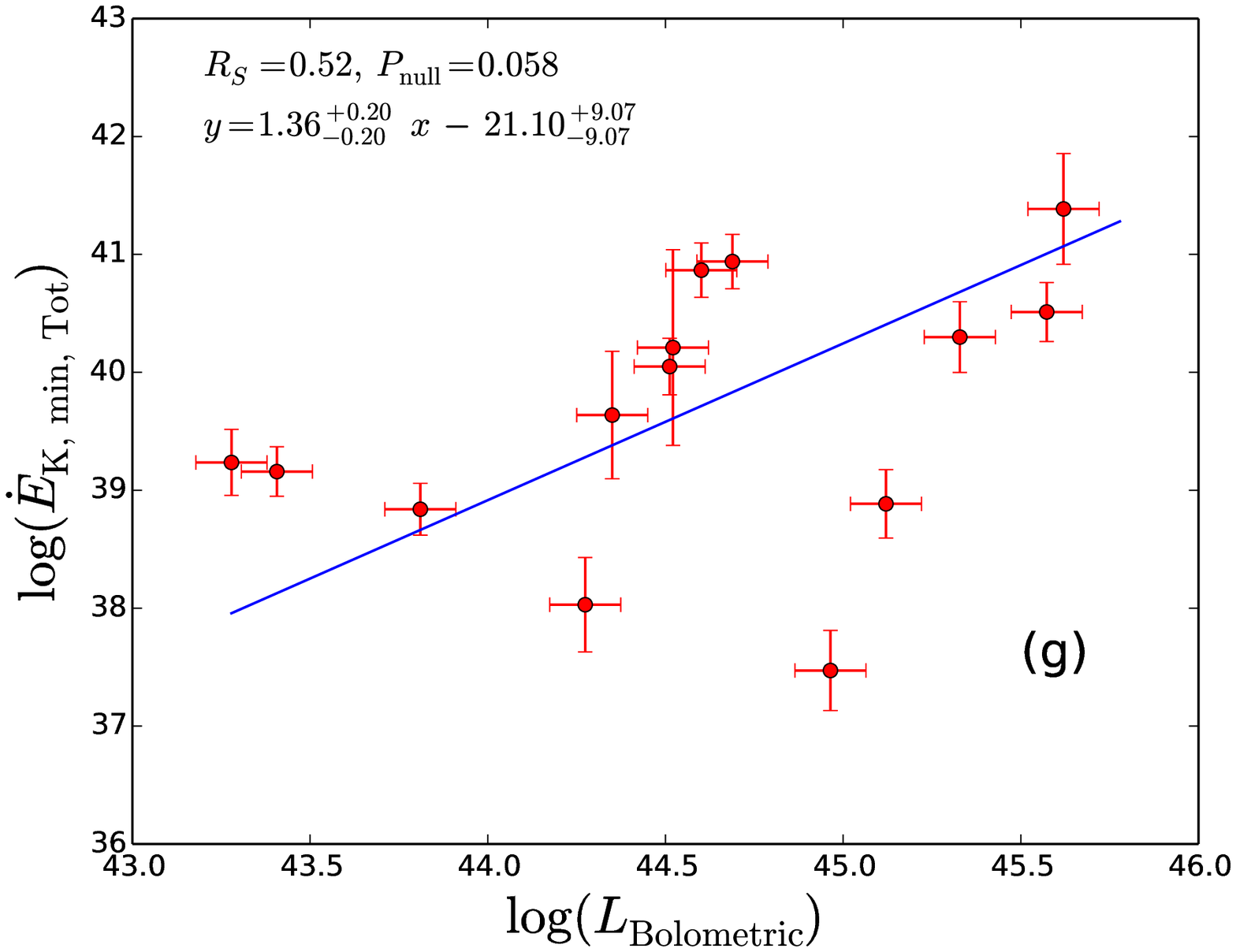}
 \includegraphics[width=7.5cm,angle=0]{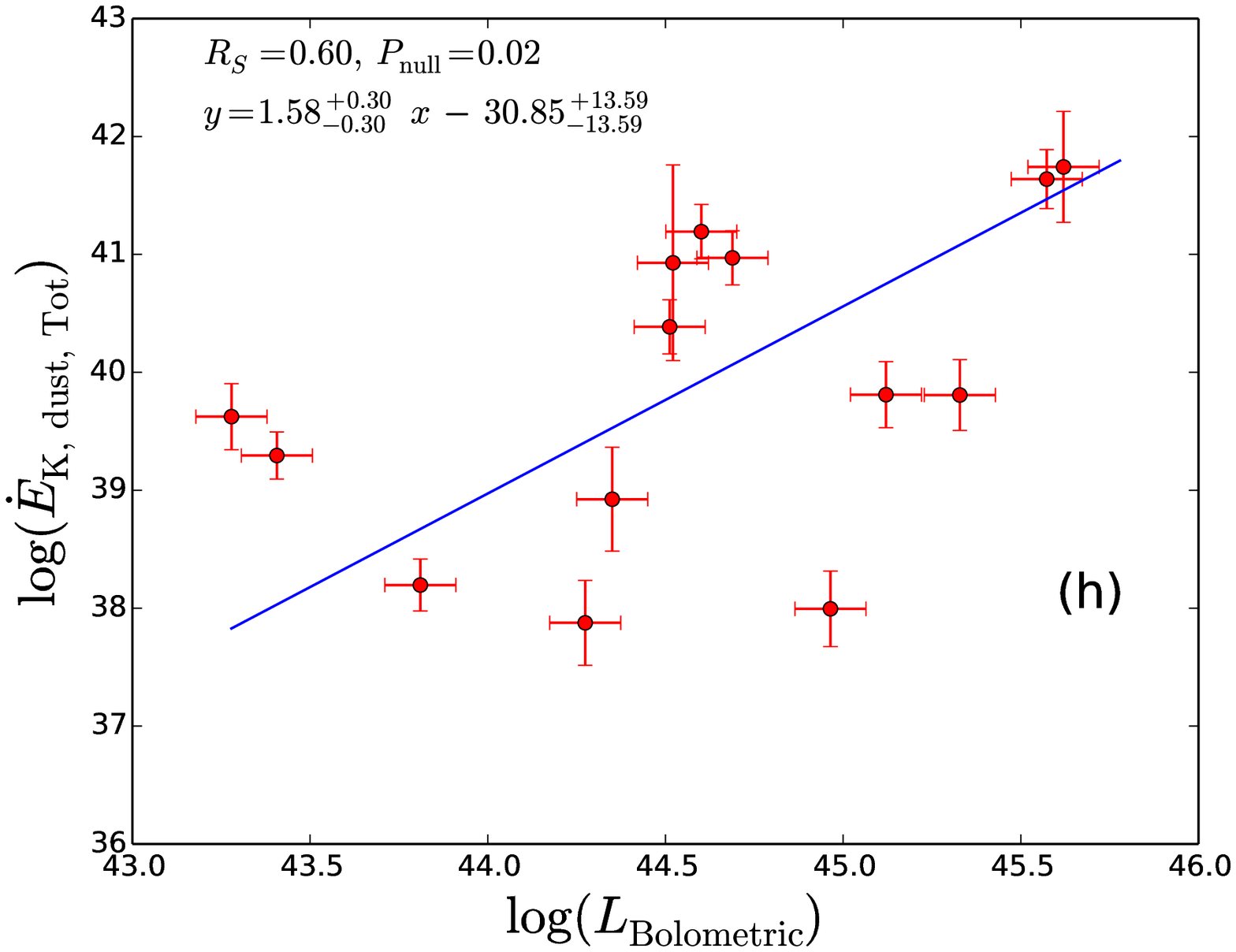}
 }

}
\caption{The correlation between the WA outflow momentum rate derived using the virial distance scale $\poutmintot$ and the dust sublimation radius $\poutdusttot$, with the source parameters. See Section \ref{subsec:acc} for details and Tables \ref{Table:corr-rmin} and \ref{Table:corr-rdust} for the correlation parameters.  }
\label{fig:min}

\end{figure*}

\begin{figure*}
  \centering

\hbox{
 \includegraphics[width=8cm,angle=0]{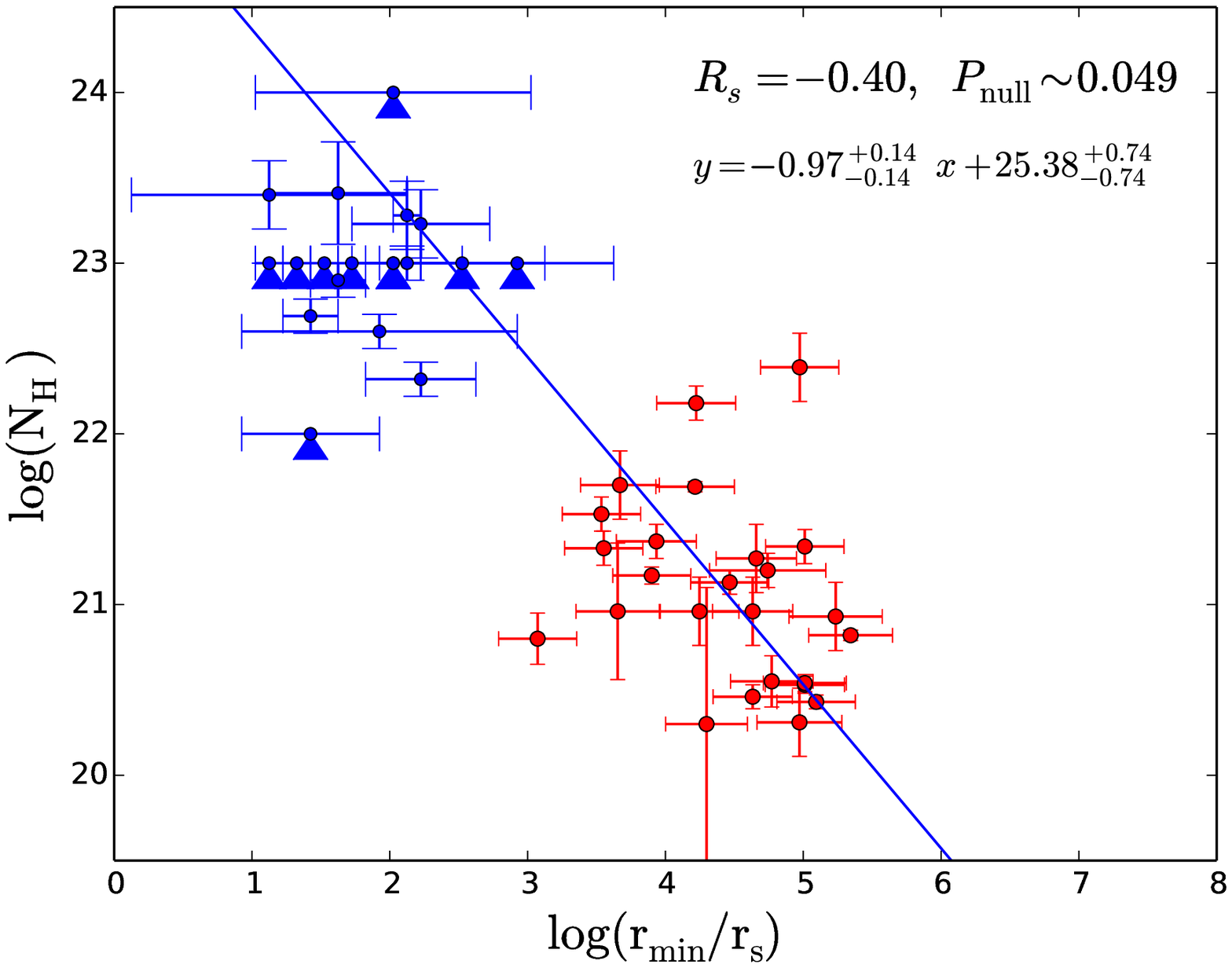}
 \includegraphics[width=8cm,angle=0]{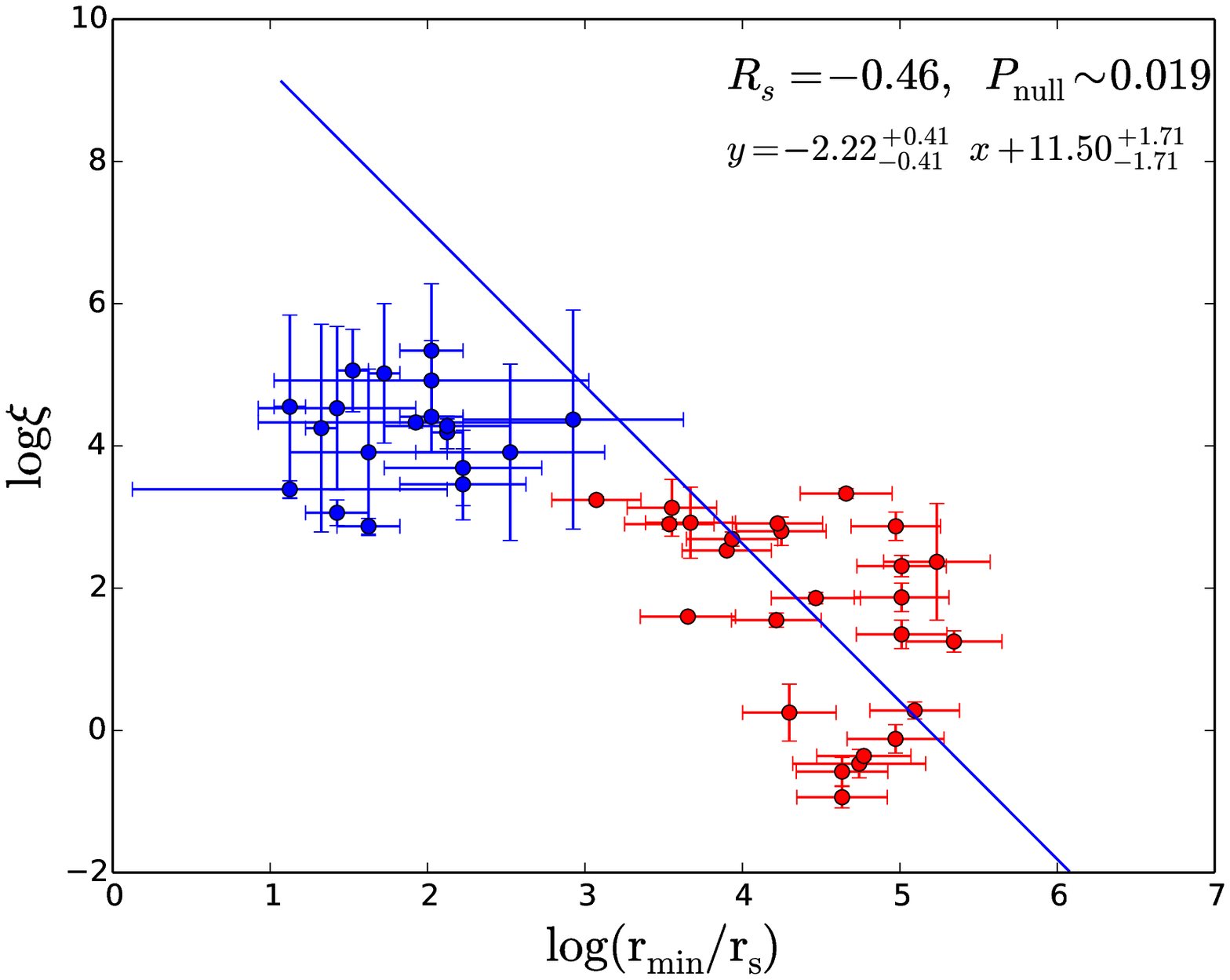}
}

\caption{The evolution of the warm absorber observables as a function of the launching radius calculated assuming virial relationship. The blue data points are the UFO from \citet{2013MNRAS.430.1102T}. The upper and lower limits of the UFO data points are shown in blue arrow-heads. The correlations are explained in Section \ref{sec:corr-analysis} }\label{fig:WA-UFO}

\end{figure*}

Panels a, b, c, d, e, f, g, and h of Figure \ref{fig:min} show the correlation between the source parameters ($\log\prad$, $\log\lbol$) with the outflow momentum rate of WA, $\log\poutmin$ and $\log\poutdust$, as well as the WA kinetic luminosity, $\log\Ekmintot$ and $\log\Ekdusttot$. The correlation and linear regression parameters are reported in Tables \ref{Table:corr-rmin} and \ref{Table:corr-rdust}. These correlations are particularly chosen to investigate the effect of source radiation on the WA outflow parameters. In an ideal situation of a purely radiatively driven wind, we would expect to see a strong positive correlation between the source luminosity and the WA momentum rates and kinetic luminosity. However, we find that the correlations in most cases show significant amount of scatter, implying that acceleration methods are more complex. Possibly the acceleration of the WA is caused by a complex combination of several mechanisms. Despite the scatter, in almost all the cases we find a positive linear trend of increase of WA outflow momentum rate and kinetic luminosity with the source luminosity.  From panels a, b, c and d, we find that the outflow momentum rate $\log\poutmin$ and $\log\poutdust$ of the WA clouds increases with both the radiation momentum rate $\log\prad$ and  bolometric luminosity $\log\lbol$ of the source. This points to the fact that a stronger source flux drives a cloud with higher momentum rate. From panels e and f we find that the total absorbed momentum rate of the WA $\log\pabs$ increases with the outflow momentum rate $\log\poutmin$ and $\log\poutdust$ pointing to the fact that the clouds with greater absorbed momentum rate have greater total outflow momentum rate. From panels g and h we find that the kinetic energy of the outflows increase with the increasing bolometric luminosity of the source. All of these correlations suggest that although the overall WA driving mechanism is complex, the incident radiation from the source plays a crucial role in accelerating the WA outflows. In other words, the WA clouds have a strong radiatively driven component.

The radiative pressure by the photons are imparted on the clouds by absorption and scattering. The UV and soft X-ray photons play the most important role in imparting momentum due to absorption, while the hard X-ray photons impart momentum to the cloud via Thomson scattering. The latter is a much weaker process in comparison to line driven force \citep{2004ApJ...615L..13E,2015MNRAS.451.2991G}. However, the question arises, how a BLR cloud located very near to the central source with strong X-ray irradiation still has enough bound electrons to be radiatively driven? This scenario has been extensively studied by \citet{2003ApJ...582...69P,2004ApJ...616..688P}. They suggest that if an already launched cloud moves up in the atmosphere of the AGN, its density falls and its temperature increases, and it gradually becomes fully ionised, and at that stage if it fails to attain the escape velocity, it falls back to the AGN disk. These ``failed wind" clouds could have gained enough opacity to shield the other clouds from the X-ray ionising photons, thereby helping them to flow out. A limitation in this `failed wind' scenario, is that, if the infalling winds screen the other parts of the wind from X-rays, we would expect to see redshifted X-ray absorption lines from these winds. But except for two cases in the WAX sample, where the warm absorber velocities are consistent with zero, all other WA are outflowing with respect to the systemic velocity. We find no trace of these screening `failed winds'.


\begin{figure*} 
  \centering
{
\hbox{
 \includegraphics[width=8cm,angle=0]{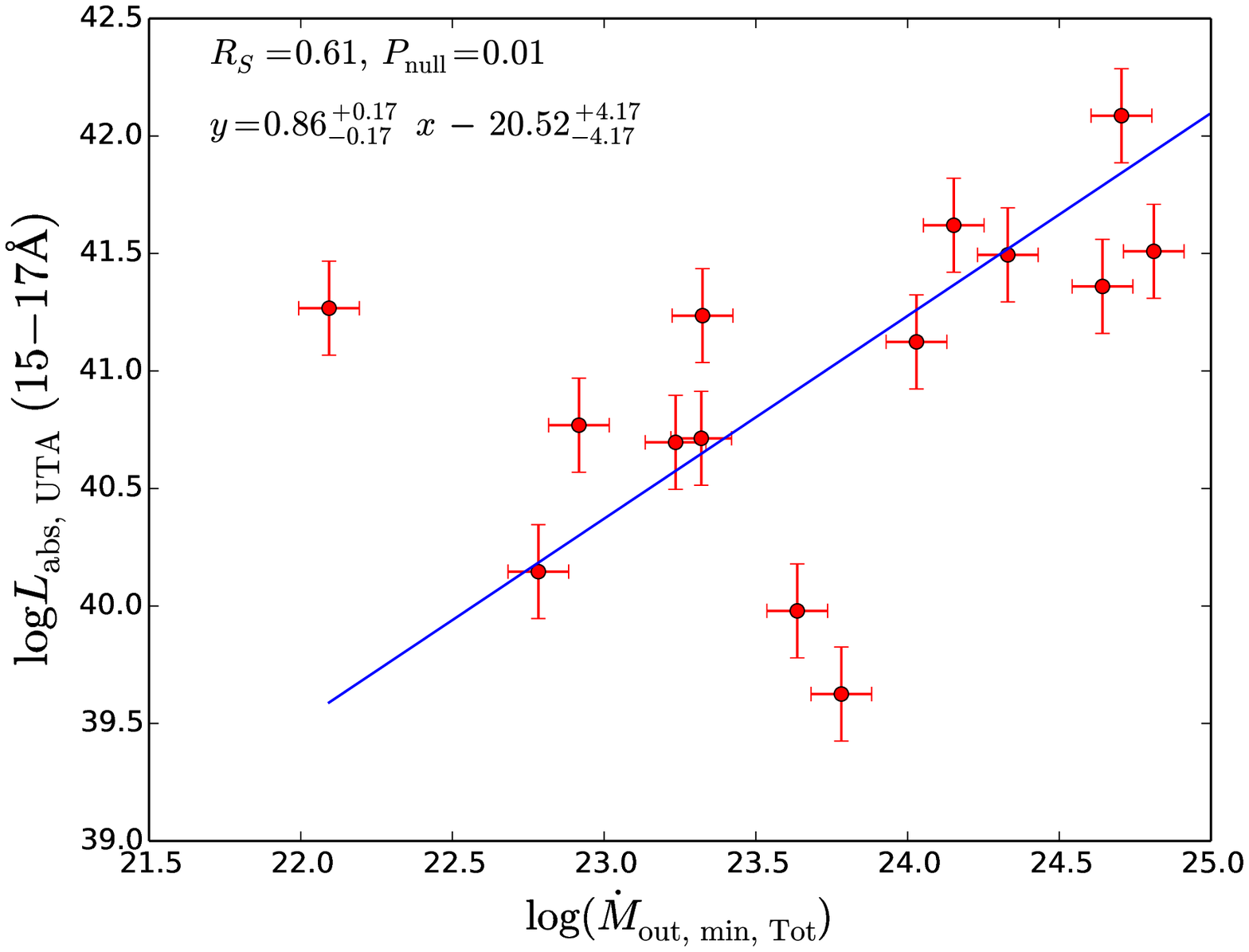}
 \includegraphics[width=8cm,angle=0]{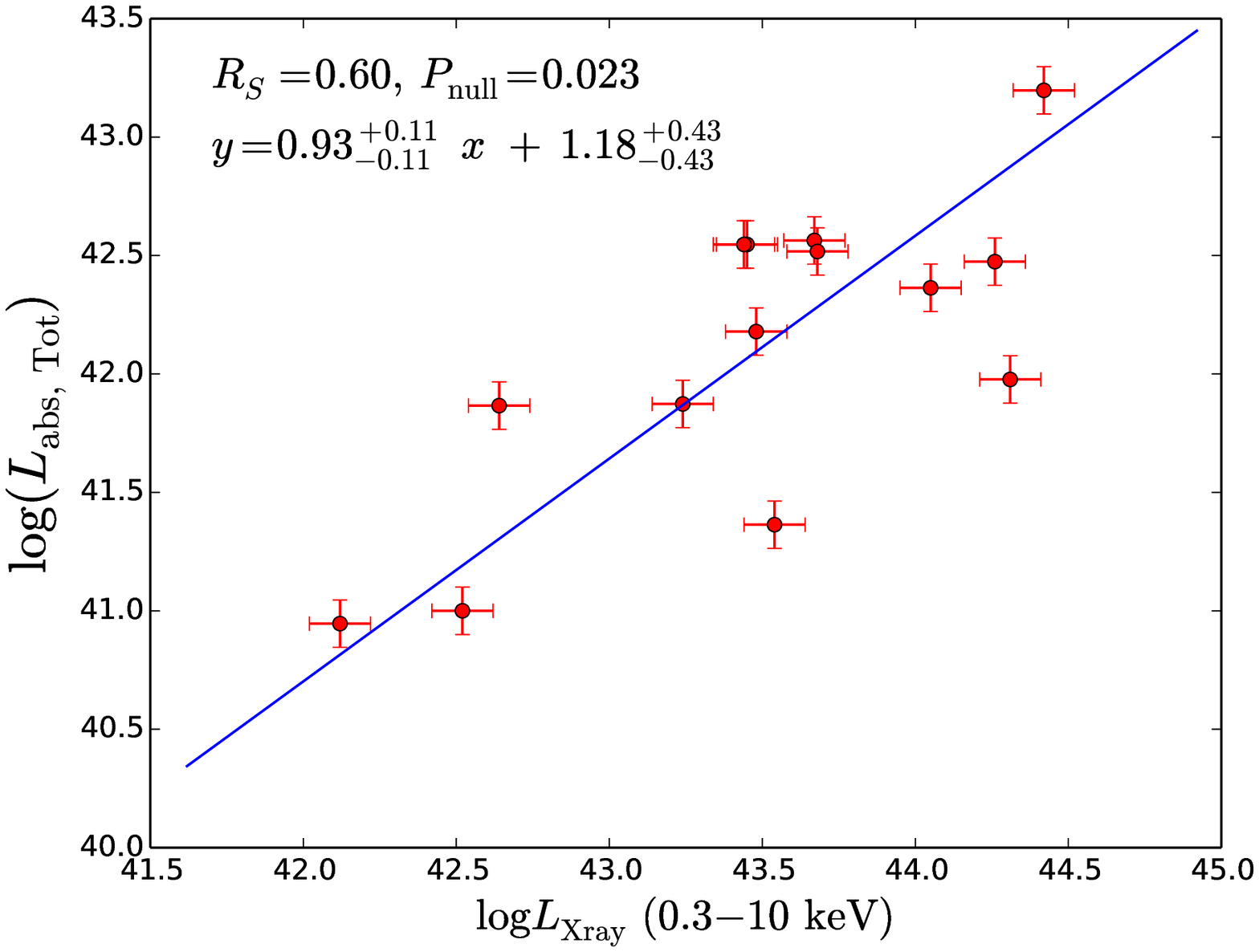}
}
}
\caption{{\it LEFT:}The correlation between the outflow mass outflow rate vs absorbed luminosity by the UTA. {\it RIGHT:} The correlation between the X-ray luminosity and the total warm absorbed luminosity for each source.}
\label{Fig:Labs_Moutmin}

\end{figure*}

\begin{figure*} 
  \centering
{

 \includegraphics[width=10.5cm,angle=90]{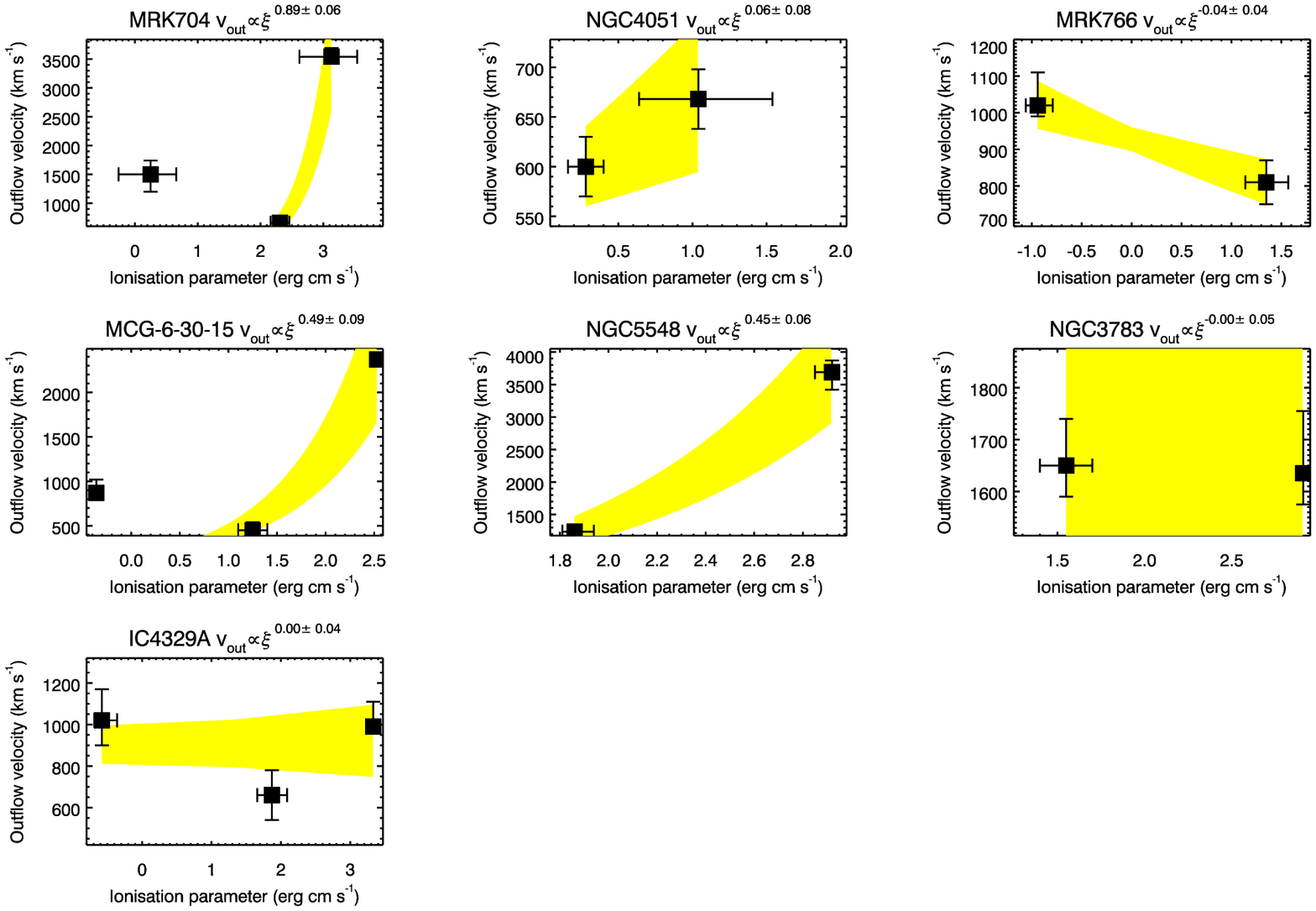}

}
\caption{The individual plots for $\log\xi$ vs $\log\vout$ for the sources with 2 or more components of WA in the WAX sample. The {\it title} of each plot reports the results of the best-fit with a function $v_{out}$~=~A$\times$$\xi^{a}$. The yellow boxes represent the envelope of the best-fit functions corresponding to a $1-\sigma$ error on the best fit parameters.}
\label{fig:logxi-logv}

\end{figure*}


Improved simulations of radiatively driven winds in AGN by  \citet{2010MNRAS.408.1396S} showed similar results that these winds could produce the higher energy absorption features ($2-10\kev$) of the spectrum of the quasar PG1211+143, but fail to produce the lower ionisation features Ne~{\sc ix} or O~{\sc vii} in the soft X-rays. This is due to the fact that the huge X-ray flux from a high Eddington ratio source has fully ionised the cloud. A recent theoretical study by \citet{2014ApJ...789...19H} found that the purely radiatively driven winds are more ionised than that was predicted by \citet{2004ApJ...616..688P}, and as such these winds will have no UV or soft X-ray spectral lines. So, in effect, the purely radiatively driven winds scenario is efficient only if there is X-ray shielding.


If indeed there is some form of X-ray shielding, we should be able to detect that in the X-ray absorption spectrum. We find an interesting correlation in Figure \ref{Fig:Labs_Moutmin}, right panel. The ratio of the total absorbed X-ray luminosity by the WA clouds $\labs$ to that of the incident X-ray luminosity ($0.3-10\kev$) is a constant at $\labs/\lxray\sim 5\%$ across the $\lxray$ values. The slope of the linear regression between $\log\lxray$ and $\log\labs$ is $\sim 1$. { The ratio of the quantity $\labstot/\lxray=10^{40.25}/10^{41.5}=0.056$ at the point where the best fit linear regression line intersects the Y-axis in the figure, implies a $\sim 5\%$ absorption.} Perhaps the WA are screening the outflowing clouds with lower ionisation states located further out from the central source radiation. The scatter in the correlation may be the result of the different viewing angles for different sources. 


{ The other possible radiative acceleration mechanism is Thomson scattering, which is particularly relevant for highly ionised clouds whose atoms have been stripped off their bound electrons. \citet{2003MNRAS.346.1025P,2010MNRAS.402.1516K} have extensively studied the case of an electron scattered wind and conjectured that in such a scenario one would expect the momentum outflow rate $\pout$ of the cloud to be proportional to the incident radiation field $\prad$, i.e., a correlation between $\log\pout$ and $\log\prad$ would yield a linear regression slope of 1.  From Fig. \ref{fig:min}, panels a and b, we find that the slopes of the linear regression are $1.06\pm0.10$ and $1.20\pm 0.19$ for $\log\poutmin$ vs $\log\prad$ and $\log\poutdust$ vs $\log\prad$ respectively. This is similar to that expected for a Thomson scattered cloud. For a Thomson scattered wind, $\pout$ and $\prad$ is predicted to be of similar magnitude. However, for the WA we find that $\pout$ is an order of magnitude lower than $\prad$. We conclude that in the WA clouds, Thomson scattering may not contribute towards accelerating the cloud. It is to be noted that this process is predominant for highly ionised, high column density clouds facing a high Eddington ratio ($\lambda=\lbol/\ledd$) source. This is the typical scenario for the ultra-fast-outflows as discussed by \citet{2013MNRAS.430.1102T} and \citet{2015MNRAS.451.4169G}.}

\citet{2008MNRAS.385L..43F,2007MNRAS.380.1172H} have shown how the presence of dust enhances the radiative thrust on the winds due to the ionising continuum. Even for sub-Eddington luminous sources, the clouds can get a thrust equivalent to sources emitting at an Eddington rate. In Sect. \ref{subsec:origin} we find that the WA can likely originate from the dust sublimation radius in the outer parts of the accretion disk or the dusty torus. This points to the possibility of presence of dust in the WA for the clouds with lower ionisation parameter $\sim \log\xi<1$. { In a dust driven scenario one would expect to get a negative slope of $\log\xi$ vs $\log\vout$ correlation, as the lower ionisation states will be driven faster. On the other hand in a purely radiatively driven wind scenario we would expect a positive slope with higher ionisation states outflowing faster. In L1, we find that the slope of $\log\xi$ vs $\log\vout$ correlation is nearly flat $\sim 0.12\pm 0.03$, which may point towards a combination of the effects of dust driven wind for the lower ionisation states and line driven wind for the higher ionisation states. } 


\subsubsection{Alternative scenarios}

Early studies on magneto-hydrodynamical (MHD) winds  by \citet{1982MNRAS.199..883B} demonstrated that energy and angular momentum can be effectively removed magnetically in the form of outflows from accretion disks, leading to visible signatures of collimated jets and uncollimated winds. More recent studies by \citet{2012ASPC..460..181K,2010ApJ...715..636F,2010ApJ...723L.228F,2004ApJ...615L..13E} presented two dimensional magneto-hydrodynamic winds, originating from the accretion disk winds irradiated by a central X-ray quasi stellar object (QSO). \citet{2010ApJ...715..636F,2010ApJ...723L.228F} found that for the chosen density profile $n(r)\propto r^{-1}$, the predicted outflow absorption features are in good agreement with the X-ray spectra from several AGN. In our study we find that the slope of the density profile for the WA is $\alpha=1.236\pm 0.034$, consistent with the assumption in the study by \citet{2010ApJ...715..636F} whose conclusions could be applied to the WAX sample as well. However, the authors also mention that for a MHD wind, the UV spectra needs to be very strong, with an $\alpha_{\rm OX}\ge 2$, which is not the case for the WAX sources, where $\alpha_{\rm OX}$ is given by the expression $0.385 \log\left[{\frac{f_{\nu}(2500\AA)}{f_{\nu}(2 keV)}}\right]$ \citep{1979ApJ...234L...9T}. An observable of an MHD driven wind is that the outflow velocity is proportional to the ionisation parameter of the outflowing cloud. In L1 we had studied the correlation between $\vout$ and $\xi$ for the WAX sample of warm absorbers and found that the linear regression slope $0.12\pm 0.03$ is inconsistent with the scaling law predicted for MHD winds ($\vout \propto \xi$). In Figure \ref{fig:logxi-logv} we show the linear regression results for individual sources, and we find that they have a wide range of values for different sources, from $0.00\pm 0.05$ in NGC~3783 to $0.89\pm0.06$ in MRK~704. This indicates that the driving mechansims of WA is varied and not one single effect can explain all the cases.

{ In view of the discussion in this section we suggest that the acceleration mechanism for warm absorbers is complex. Not a single physical process can capture the entire physics. However, radiative acceleration by UV and soft X-ray line absorption is a predominant mechanism. The contributions from MHD-driven mechanisms cannot be ruled out.}


\subsection{The WA feedback to galaxies} \label{sec:feedback}
\subsubsection{Caveats on $\rmax$ estimates.}\label{subsec:rmax}

 In this section we discuss the caveats and shortcomings of the estimates of $\rmax$. Firstly, $\rmax$ arises from a geometrical constraint $\Delta r = r$, where $\Delta r$ is the thickness of the cloud and $r$ is the distance of the inner surface of the WA cloud facing the ionizing radiation, where the ionisation parameter $\xi$ for the cloud is defined. The equality implies that inner surface distance of the cloud from the ionising source and the cloud thickness are equal. The inner surface distance and the thickness of the cloud are entirely unrelated quantities. Instead the outer surface distance of the cloud ($R$) could put a limit on the thickness of the cloud ($ \Delta r \le R$). But unfortunately, we do not have a measure of the outer extent of the cloud. However, keeping in mind the outflow literature published so far, we have calculated $\rmax$ using equation \ref{equ:rmax} for every WA component for a comparison.

Secondly, from Table \ref{Table:dist} we find that at least one of the WA has $\rmax \sim 1 \mpc$ and five others have $\rmax \sim 100\kpc$. \citet{2013MNRAS.430.1102T} in a compilation of the warm absorber studies found similar $\rmax$ ranging upto $10^{25}\cm \sim 10\mpc$, using equation \ref{equ:rmax}. { \citet{2012ApJ...753...75C} have also calculated $\rmax$ of the order of $10^{23}-10^{26}\cm \sim 100\kpc-100\mpc$ for WA in several sources}, using equation \ref{equ:rmax}. These values imply that the WA can be found much beyond the host galaxies and may be sometimes in the inter galactic medium (IGM). Most of the robust WA distance estimates from variability studies put them between $0.001\pc$ to few tens of $\pc$ (see Table \ref{Table:WA-var}), which in turn questions the validity of equation \ref{equ:rmax}.

Keeping in mind these caveats, we estimate the outflow parameters corresponding to $\rmax$ and discuss the results in the sections below.

\subsubsection{WA feedback}

A measure of the WA feedback to host galaxies can be obtained by studying the WA mass outflow rate $\mout$ in comparison to the SMBH mass accretion rate $\macc$ and the kinetic luminosity of the outflow $\Ekout$ in comparison to the bolometric luminosity $\lbol$ of the source. Some of the most interesting questions regarding the AGN-galaxy co-evolution can be answered by observational studies of outflow impact on the host galaxies. For e.g., chemical enrichment of the host galaxy, the relation between the mass of SMBH and the stellar velocity dispersion ($M-\sigma$ relation), the formation of large scale structures in the early universe and the relation between SMBH and galactic bulge growth. \citet{2005Natur.433..604D,2010MNRAS.401....7H} have demonstrated that an outflow with $\Ek/\lbol$ as low as $\sim 0.5\%$ can give sufficient feedback to the host galaxy to stop the cooling flows in the host galaxy which supply the raw material to the accretion disk. Seyferts are primarily hosted in spiral galaxies, for which this percentage is much lower, as low as $0.1\%$ \citep{2011MNRAS.411..349G,2012MNRAS.424..190G} to affect the host galaxy, as against the more compact elliptical galaxies.

\begin{figure*} 
  \centering

 \includegraphics[width=10cm,angle=0]{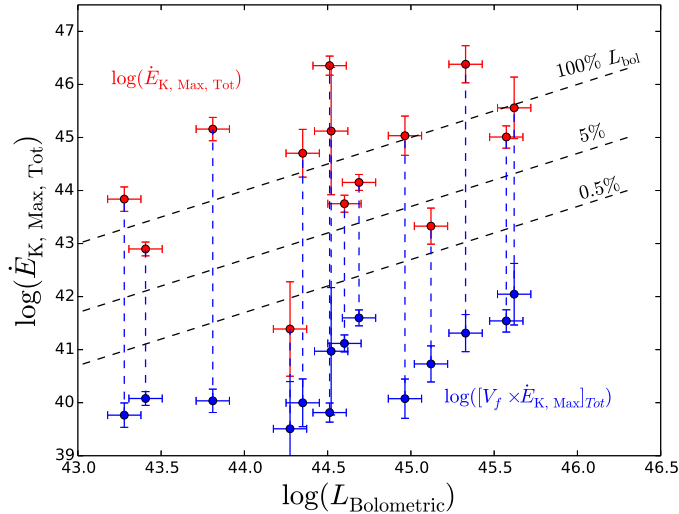}

\caption{The total maximum kinetic luminosity of the WA for each source plotted against the bolometric luminosity. We find that for a unit volume filling factor $V_f$, the kinetic luminosity can exceed $\lbol$ by nearly an order of magnitude (upper red dots). However the uncertainty in the measurement of $V_f$ leads to a large uncertainty in the calculation of $\Ek$. Following \citet{2005A&A...431..111B} we estimated $V_f$ in Section \ref{subsec:vol} and reported them in Table \ref{Table:Ek}. We note that if we consider the volume filling factor, the kinetic outflow rates for the WA are lower than $0.5\%$ of the bolometric luminosity (lower blue dots), which implies that the WA may not have sufficient feedback impact on its surroundings. }   \label{Fig:Ek}
\end{figure*}

 From Figure \ref{Fig:Ek} and Table \ref{Table:Ek} we find that the WA maximum kinetic energy can exceed the bolometric luminosity of the AGN, sometimes even by an order of magnitude, for unit volume filling factor $V_f$ and assuming $\rmax$ estimate in equation \ref{equ:rmax}. However, if we multiply $\Ekmax$ of each of the WA component by the respective $V_f$ (see Table \ref{Table:Ek}), we obtain the blue points in Fig. \ref{Fig:Ek}. We find that after taking into account the volume filling factor, the kinetic luminosity of the WA are $<0.5\% \lbol$. There is however, a large uncertainty in the estimate of the filling factor.



\begin{figure*}
 \centering
 \hbox{
 \includegraphics[width=8cm,angle=0]{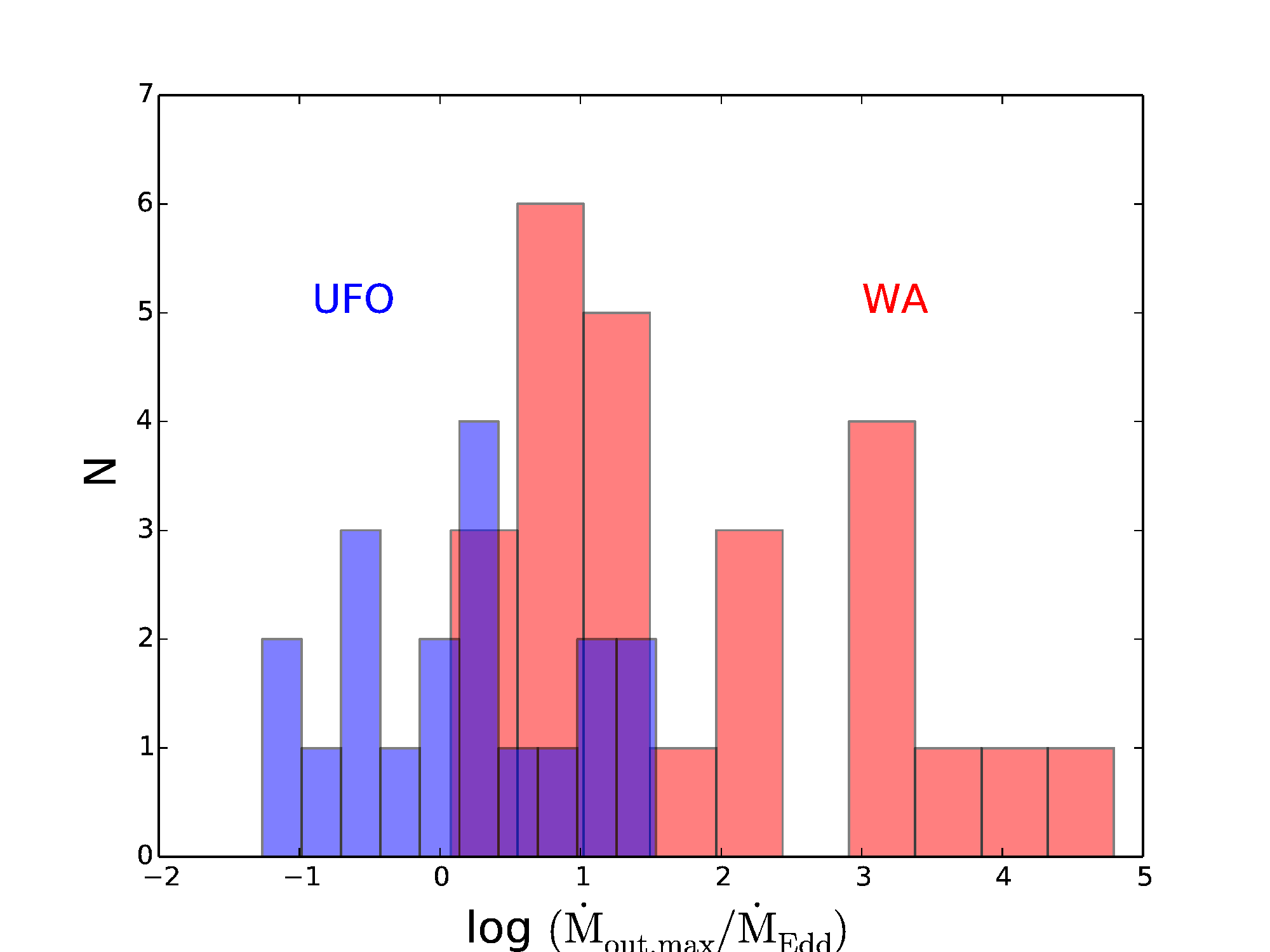}
\includegraphics[width=8cm,angle=0]{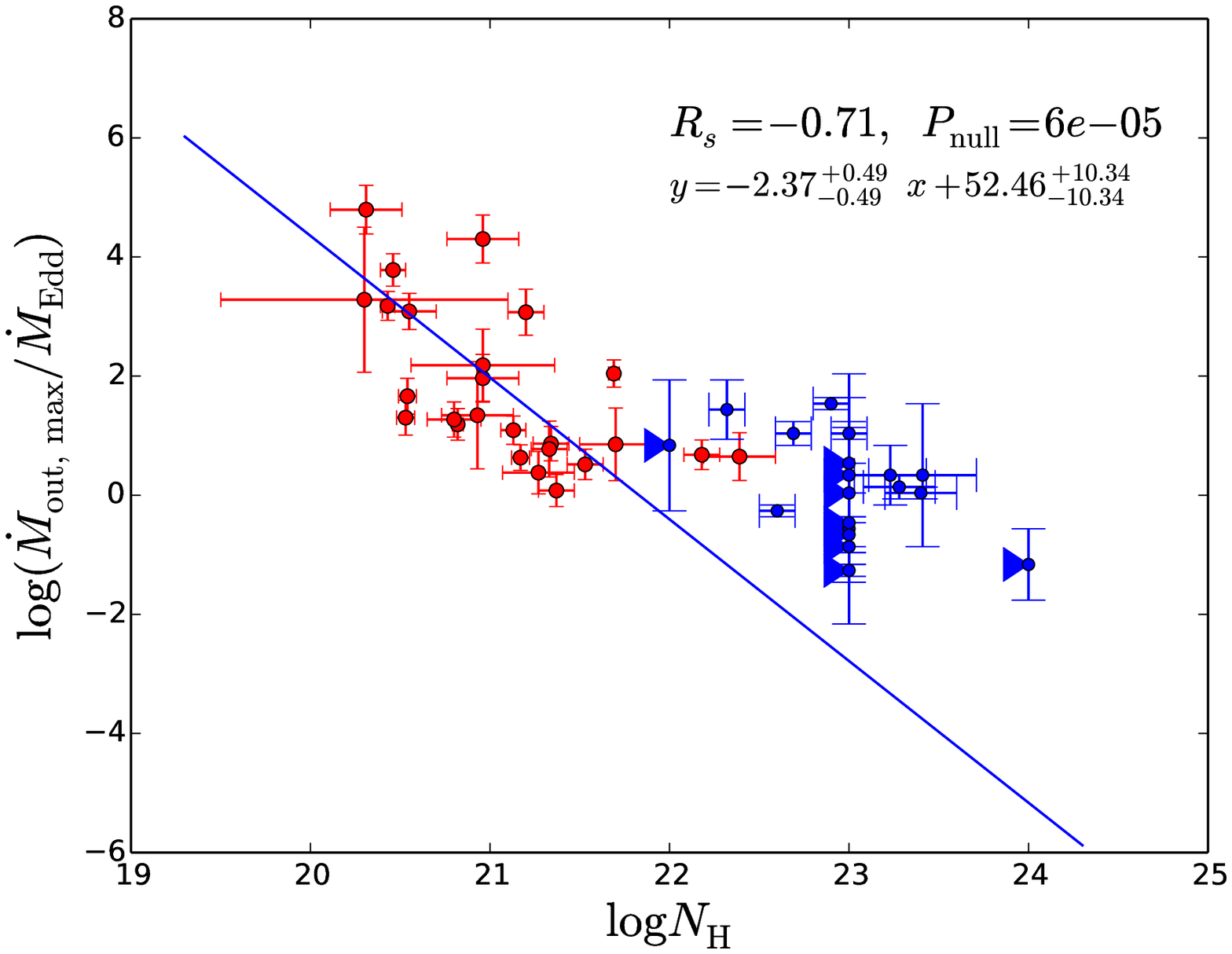}

}
 
\caption{{\it LEFT:} Histograms showing the maximum mass outflow rates for WA (in red) for unit volume filling factor, scaled with respect to the Eddington rate. The blue histograms are the mass outflow rates for UFO. {\it RIGHT:} The scaled mass outflow rate of the WA and the UFO (in blue). }

\label{fig:Moutmax}
\end{figure*}

 From Figure \ref{fig:Moutmax}, left panel, we find that the distribution of the maximum mass outflow rates of the warm asorbers (red histograms), scaled with respect to the Eddington mass accretion rate, exceed the latter by several orders of magnitude ($\sim 10^5$), { for unit volume filling factor}. Previous studies by \citet{2007MNRAS.379.1359M} and \citet{2012ApJ...753...75C} have also found that $\mout/\medd$ is of the factor $10-1000$ for unit volume filling factor. { In a study on the WA of the source MR~2251-178, \citet{2011MNRAS.414.3307G} have found that a clumping factor of $\sim 10^{-3}$ is required for the mass outflow rates of the WA not to greatly exceed the accretion rate. The large mass outflow rate as compared to the mass accretion rate indicates that most of the matter which is approaching the black hole is being expelled and only a small fraction is reaching the black hole.} 

From Figure \ref{fig:Moutmax} right panel we find that the more massive outflows are the ones which have lower column densities ($\nh$) or alternatively lower ionisation parameter $\xi$, { for unit volume filling factor}. Previous studies suggested that Fe M shell unresolved transition array (UTA) absorption could carry away the largest mass outflows \citep{2005ApJ...632..788H,2007MNRAS.379.1359M}. We calculated the absorbed luminosity for each warm absorbed source in the wavelength range $15-17\rm \AA$ which we assume to be the representative of the UTA absorption. Fig. \ref{Fig:Labs_Moutmin} left panel,  show that there is a correlation between the UTA absorbed luminosity with the total mass outflow rate. The linear trend suggests that higher UTA absorption leads to higher mass outflow rates. This indicates that the UTA absorbers may be the dominant mass outflow component.


\subsection{The WA and the UFO: Are they similar outflows?}

There has been long standing question as to whether the warm absorbers and the UFOs are the same type of outflows. Our analysis in L1 had shown that the WA and the UFO do not lie on the same correlation slope in the parameter spaces of $\xi$, $\nh$ and $\vout$, and hence most probably not the same type of outflow. In principle, WA and UFOs denote two very different types of cloud with distinctly different $\xi$, $\nh$ and $\vout$ distributions. Below we discuss the relation between WA and the UFO in the light of the dynamical calculations done in this work.

 Fig. \ref{fig:WA-UFO} left panel shows the plot of the column density $\nh$ vs the launching radius $\rmin$ for both the WA (in red) and the UFO (in blue). The correlation slope between the WA parameters encompasses the UFOs, implying a continuous column density distribution over the WA and UFO geometrical distances. On the other hand, the right panel of Fig. \ref{fig:WA-UFO} shows the correlation between the launch radius and the ionisation parameter, with a break in the continuity of the ionisation parameter of the WA and UFO. The distance at which this break occurs is $\sim 10^3\rs$, and as discussed in Section \ref{subsec:acc}, this could be due to shock entrainment of the UFOs with the ISM, giving rise to the WA.

The properties of the UFO and the WA are quite different. The UFOs originate at a much nearer radius ($\sim 10 \rs$) and are already highly ionised by the incoming continuum radiation, which incapacitates them from being line driven. They are therefore primarily driven by Thomson scattering. On the other hand, the WA are primarily radiatively driven with dust contributing to its opacity. \citet{2013MNRAS.430.1102T} has unified the UFO and the WA clouds by describing the WA clouds to be the cooler and slower form of UFOs, located further out in the AGN. They mention that the UFO and the WA are a part of the single large stratified outflow. In our study we find that the WA likely originate from the inner regions of the dusty torus, while the UFOs originate from accretion disk. Thus WA are a different type of wind compared to the UFO. However, considering the shock driven theory of WA origin by \citet{2013MNRAS.433.1369P}, it could be possible that the WA originate due to shock impact of the UFO on the neutral and dusty torus or the ISM.


\section{Conclusions}

We have carried out a comprehensive study on the warm absorber dynamics and its effect on the environments of the host galaxy. The main conclusions of the paper are as follows: \\

\begin{itemize}

\item We find that the linear regression slope between the absorbed ionizing luminosity $\log\labstot$ vs X-ray luminosity $\log\lxray$ is $0.93\pm 0.11$ indicating a constant fractional absorption ($\sim 5\%$) independent of X-ray flux. 

\item We estimate WA launching radius using the virial argument, and the dust sublimation radius (following \citet{2011A&A...525L...8C}). The distribution of launching radii is broad, ranging between 10$^3$ and 10$^6 \rs$. We find that the WA possibly originates as a result of photo-ionised evaporation from the inner edge of the torus (Torus wind), or as a shock driven wind generated as a result of the impact of the UFO on the ISM or the torus, at a typical distance of $10^3 \rs$.

\item The WA density profile estimated from the linear regression slope of $\xi$ and $\nh$ in the WAX sample is $\alpha=1.236\pm 0.034$, assuming that the different WA across the different sources are manifestations of the different stages of the same absorber. This result agrees with the detailed study on a few sources exhibiting multi-component WA outflows, validating our assumption.

\item The acceleration mechansim of the WA is complex and neither radiatively driven wind nor MHD driven wind scenario alone can describe the outflow acceleration. However, we find that the radiative forces play a significant role in driving the outflows through soft X-ray absorption lines, { as well as dust opacity.}

\item To estimate the feedback of warm outflows onto the host halaxy, accurate estimates of the volume filling factor as well as of the outflow's radial extension are required. The best available estimates of these quantities suggest that the WA components in the WAX sample do not generally exert a significant feed-back. However, given the large uncertainties on these geometrical factors, at least some WA may exert sufficient feed-back to contribute to the cosmological evolution and to the chemical enrichment of the host galaxy \citep{2012ApJ...753...75C}.

\item { We find that { for unit volume filling factor,} the maximum mass outflow rates are several orders of magnitude higher than the mass accretion rates, consistent with previous estimates.}

\item The lowest ionisation states carry the maximum mass outflow. The sources with the strongest absorption in the Fe-M UTA wavelength range exhibit the largest mass outflow rates.

\item  WA and UFOs correspond to two different types of outflow with distinctly different ionisation states, column densities and outflow velocities. The launching radii of the UFO and the WA are separated at about $10^3\rs$. For every source where both WA and UFO have been detected, we have compared the highest ionisation parameter of WA with the highest ionisation parameter of UFO and their corresponding outflow velocities. We find that the ratio of the maximum UFO velocity with the highest WA velocity for the sources that exhibit both UFO and WA is $\sim 10-30$ while the drop in the ionisation parameter is larger $\sim 10-10^3$.

\end{itemize}

$Acknowledgements:$  The author SL is grateful to Stuart Sim, Gary Ferland, Chris Reynolds, Chris Done, Andy Fabian, Poshak Gandhi and Sebastian Hoenig for insightful discussion on various aspects of this paper. This research has made use of the NASA/IPAC Extragalactic Database (NED) which is operated by the Jet Propulsion Laboratory, California Institute of Technology, under contract with the National Aeronautics and Space Administration. Author SL is grateful to the science and technology facilities council (STFC), U.K.


\begin{table*}

{\footnotesize
\centering
  \caption{The linear regression analysis results using the virial radius $\rmin$ for derived outflow parameters ($y=a\,x+b$) following the procedure enumerated in Section \ref{sec-corr}. All quantities are in logarithmic units. See Figure \ref{fig:min}. \label{Table:corr-source}}
  \begin{tabular}{llllllllllll} \hline\hline 

 Serial & $x$ 			 & $y$ 		& $a$ 		&  \hspace{1cm}Dev$(a)$ & \hspace{1cm} $b$ & \hspace{1cm} Dev$(b)$ & \hspace{1cm}$R_S$ &\hspace{1cm} $P_{null}$\\ \hline \\

1.	&$\lxray$ 	& 	$\labstot$	&$0.93$ 	&\hspace{1cm} $0.11$ & \hspace{1cm}$1.18$  & \hspace{1cm}$0.43$ & \hspace{1cm}$0.60$ & \hspace{1cm} $0.023$\\ 

2.	&$\rmin/\rs $	&$ \nh$			&$-0.97$ 	&\hspace{1cm} $0.15$ & \hspace{1cm}$25.38$  & \hspace{1cm}$0.76$ & \hspace{1cm}$-0.40$ & \hspace{1cm} $0.049$\\ 

3.	&$\rmin/\rs $	&$ \xi$			&$-2.22$ 	&\hspace{1cm} $0.41$ & \hspace{1cm}$11.50$  & \hspace{1cm}$1.71$ & \hspace{1cm}$-0.46$ & \hspace{1cm} $0.019$\\ 

4.	&$ \nh $		&$\mout/\medd$	&$-2.37$ 	&\hspace{1cm} $0.49$ & \hspace{1cm}$52.46$  & \hspace{1cm}$10.37$ & \hspace{1cm}$-0.71$ & \hspace{1cm} $\rm 6.19e-05$ \\

5.	&{ $ \moutmintot $}		&{$\labsuta$}	&$0.86$ 	&\hspace{1cm} $0.17$ & \hspace{1cm}$-20.52$  & \hspace{1cm}$4.17$ & \hspace{1cm}$0.61$ & \hspace{1cm} $\rm 0.01$ \\\hline \hline

\end{tabular}  

\footnotetext{1}{ $R_S$ stands for the Spearman rank correlation coefficient.}

}
\end{table*}

\begin{table*}

{\footnotesize
\centering
  \caption{The linear regression analysis results using the virial radius $\rmin$ for derived outflow parameters ($y=a\,x+b$) following the procedure enumerated in Section \ref{sec-corr}. All quantities are in logarithmic units. See Figure \ref{fig:min}. \label{Table:corr-rmin}}
  \begin{tabular}{llllllllllll} \hline\hline 

 Serial & $x$ 			 & $y$ 		& $a$ 		&  \hspace{1cm}Dev$(a)$ & \hspace{1cm} $b$ & \hspace{1cm} Dev$(b)$ & \hspace{1cm}$R_S$ &\hspace{1cm} $P_{null}$\\ \hline \\

1.	&$\prad$		&	$\pouttot$ 	&$1.06$ 	&\hspace{1cm} $0.10$ & \hspace{1cm}$-4.07$  & \hspace{1cm}$3.53$ & \hspace{1cm}$0.34$ & \hspace{1cm} $0.23$\\ 

2.	&$\pouttot$	&	$\pabstot$  &$0.77$ 	&\hspace{1cm} $0.17$ & \hspace{1cm}$-6.93$  & \hspace{1cm}$5.43$ & \hspace{1cm}$0.56$ & \hspace{1cm} $0.03$\\ 

3.	&$\lbol$		& $\Ektot$  &$1.36$	&\hspace{1cm} $0.20$ & \hspace{1cm}$21.10$  & \hspace{1cm}$9.07$ & \hspace{1cm}$0.52$ & \hspace{1cm} $0.058$\\ 

4.	&$\lbol$		& $\pouttot$  &$1.11$	&\hspace{1cm} $0.17$ & \hspace{1cm}$-17.59$  & \hspace{1cm}$7.58$ & \hspace{1cm}$0.45$ & \hspace{1cm} $0.10$\\

\hline \hline
\end{tabular} 

\footnotetext{1}{ $R_S$ stands for the Spearman rank correlation coefficient.}

}
\end{table*}


\begin{table*}

{\footnotesize
\centering
  \caption{The linear regression analysis using $\rdust$ for derived outflow parameters ($y=a\,x+b$) following the procedure enumerated in Section \ref{sec-corr}. All quantities are in logarithmic units. \label{Table:corr-rdust}}
  \begin{tabular}{llllllllllll} \hline\hline 

 Serial & $x$ 			 & $y$ 		& $a$ 		&  \hspace{1cm}Dev$(a)$ & \hspace{1cm} $b$ & \hspace{1cm} Dev$(b)$ & \hspace{1cm}$R_S$ &\hspace{1cm} $P_{null}$\\ \hline \\

1.	&$\prad$		&	$\pouttot$ 	&$1.20$ 	&\hspace{1cm} $0.19$ & \hspace{1cm}$-8.90$  & \hspace{1cm}$6.55$ & \hspace{1cm}$0.47$ & \hspace{1cm} $0.09$\\ 

2.	&$\pouttot$	&	$\pabstot$  &$0.81$ 	&\hspace{1cm} $0.15$ & \hspace{1cm}$-5.68$  & \hspace{1cm}$4.96$ & \hspace{1cm}$0.37$ & \hspace{1cm} $0.19$\\ 

3.	&$\lbol$		& $\Ektot$ &$1.58$ 	&\hspace{1cm} $0.30$ & \hspace{1cm}$30.85$  & \hspace{1cm}$13.59$ & \hspace{1cm}$0.60$ & \hspace{1cm} $0.02$\\

4.	&$\lbol$		& $\pouttot$ &$1.28$ 	&\hspace{1cm} $0.23$ & \hspace{1cm}$24.92$  & \hspace{1cm}$10.54$ & \hspace{1cm}$0.59$ & \hspace{1cm} $0.02$\\

 \hline \hline

\end{tabular} 

\footnotetext{1}{ $R_S$ stands for the Spearman rank correlation coefficient.}

}
\end{table*}


\begin{table*}

{\footnotesize
\centering
  \caption{The linear regression analysis using { $\rmax$} for deriving outflow parameters ($y=a\,x+b$) following the procedure enumerated in Section \ref{sec-corr}. All quantities are in logarithmic units. \label{Table:corr-rmax}}
  \begin{tabular}{llllllllllll} \hline\hline 

 Serial & $x$ 			 & $y$ 		& $a$ 		&  \hspace{1cm}Dev$(a)$ & \hspace{1cm} $b$ & \hspace{1cm} Dev$(b)$ & \hspace{1cm}$R_S$ &\hspace{1cm} $P_{null}$\\ \hline \\

1.	&$\prad$		&	$\pouttot$ 	&$1.52$ 	&\hspace{1cm} $0.32$ & \hspace{1cm}$-15.19$  & \hspace{1cm}$11.19$ & \hspace{1cm}$0.47$ & \hspace{1cm} $0.087$\\ 

2.	&$\pouttot$	&	$\pabstot$  &$0.51$ 	&\hspace{1cm} $0.10$ & \hspace{1cm}$13.02$  & \hspace{1cm}$3.82$ & \hspace{1cm}$0.70$ & \hspace{1cm} $0.005$\\ 

3.	&$\lbol$		& $\Ektot$  &$1.60$ 	&\hspace{1cm} $0.33$ & \hspace{1cm}$-26.97$  & \hspace{1cm}$14.99$ & \hspace{1cm}$0.39$ & \hspace{1cm} $0.15$\\ \hline \hline
\end{tabular}  

\footnotetext{1}{ $R_S$ stands for the Spearman rank correlation coefficient.}

}
\end{table*}




\begin{table*}

{\footnotesize
\caption{The luminosities of the sources in the WAX sample.} \label{Table:lum} 
  \begin{tabular}{lllllllllllllll} \\ \hline\hline 

Serial	&Sources	&  $L_{\rm 2-10 \kev}$ \hspace{0.5cm}	& $\lbol^a$ \hspace{0.5cm} & $\lion^b$   \\ 

	&		& ($\lunit$)				&($\lunit$)			&($\lunit$) \\ \hline \\

1	&NGC~4593	& $8.22 \times 10^{42}$	& $1.88 \times 10^{44}$	& $1.37\times10^{44}$	\\

2	&MRK~704	& $2.13 \times 10^{43}$	& $3.32 \times 10^{44}$	& $2.18\times10^{44}$	\\

3	&ESO~511-G030	& $2.29 \times 10^{43}$	& $1.78 \times 10^{44}$	&  $1.23\times10^{44}$      \\

4	&NGC~7213	& $6.74 \times 10^{41}$	& $1.08 \times 10^{43}$	&  $5.74\times10^{42}$	\\

5	&AKN~564	& $1.95 \times 10^{43}$	& $9.21 \times 10^{44}$	&  $7.92\times 10^{44}$	\\

6	&MRK~110	& $8.12 \times 10^{43}$	& $3.12 \times 10^{45}$	&  $2.37\times 10^{44}$	\\

7	&ESO~198G024	& $4.75 \times 10^{43}$	& $7.03 \times 10^{44}$	&  $4.21\times 10^{44}$	\\

8	&Fairall~9	& $1.29 \times 10^{45}$	& $8.92 \times 10^{45}$	&  $5.68\times10^{45}$      \\

9	&UGC~3973	& $8.98 \times 10^{42}$	& $7.14 \times 10^{44}$	&  $2.37\times 10^{44}$	\\

10	&NGC~4051	& $4.64 \times 10^{41}$	& $2.55 \times 10^{43}$	&  $2.20\times 10^{43}$	\\

11	&MCG-2-58-22	& $1.42 \times 10^{44}$	& $1.79 \times 10^{45}$	&  $1.13\times 10^{45}$	\\

12	&NGC~7469	& $1.37 \times 10^{43}$	& $1.32 \times 10^{45}$	&  $1.06\times 10^{45}$	\\

13	&MRK~766	& $9.25 \times 10^{42}$	& $6.46 \times 10^{43}$	&  $4.96\times 10^{43}$      \\

14	&MRK~590	& $1.00 \times 10^{43}$	& $8.32 \times 10^{44}$	&  $4.15\times10^{43}$	\\

15	&IRAS~05078	& $1.65 \times 10^{43}$	& $2.24 \times 10^{44}$	&  $7.46\times10^{43}$	\\

16	&NGC~3227	& $1.79 \times 10^{42}$	& $3.41 \times 10^{43}$	&  $1.92\times10^{43}$	\\

17	&MR~2251-178	& $1.89 \times 10^{44}$	& $4.17 \times 10^{45}$	&   $1.40\times10^{45}$	\\

18	&MRK~279	& $5.54 \times 10^{43}$	& $2.06 \times 10^{45}$	&  $1.54\times10^{45}$	\\

19	&ARK~120	& $8.90 \times 10^{43}$	& $5.47 \times 10^{45}$	&  $3.49\times10^{45}$	\\

20	&MCG+8-11-11	& $4.01 \times 10^{43}$	& $6.77 \times 10^{44}$	&  $3.23\times10^{44}$	\\

21	&MCG-6-30-15	& $7.06 \times 10^{42}$	& $1.90 \times 10^{43}$	&  $1.41\times10^{43}$      \\

22	&MRK~509	& $8.82 \times 10^{43}$	& $3.74 \times 10^{45}$	&  $2.42\times10^{45}$	\\

23	&NGC~3516	& $2.65 \times 10^{42}$	& $3.25 \times 10^{44}$	&  $2.97\times10^{43}$	\\

24	&NGC~5548	& $2.65 \times 10^{43}$	& $3.99 \times 10^{44}$	&  $2.10\times10^{44}$	\\

25	&NGC~3783	& $1.32 \times 10^{43}$	& $4.88 \times 10^{44}$	&  $3.45\times10^{44}$	\\

26	&IC~4329A	& $5.73 \times 10^{43}$	& $2.13 \times 10^{45}$	&  $1.88\times10^{45}$	\\

\hline \hline
\end{tabular} \\ 

{$^a$ The bolometric luminosity $\lbol$ is calculated in the energy band $1\ev-250\kev$ using the best fit SEDs obtained in our earlier study L1. \\}
{$^b$ The ionising luminosity $\lion$ is calculated in the energy band $13.6\ev-13.6\kev$ (1-1000 Ryd) using the best fit SEDs obtained in our earlier study L1 }
}

\end{table*}


\begin{table*}

{\footnotesize
\centering
\caption{The minimum, dust and maximum distances of the warm absorber components of the WAX sample of sources.} \label{Table:dist} 
  \begin{tabular}{lllllllllllllll} \hline\hline 

Sources &$\labs$	& $\log\rmin$ 		&$\log\rdust$		&$\log\rmax^1$	&$\log\rs$		&$\log\rtorus$	\\ 
	&$\lunit$	& (cm)			& (cm)			& (cm)		& (cm)			& (cm) \\ \hline \\ 

NGC~4593 	&$4.62e+41$ 	&$17.48\pm 0.27$	&$17.32\pm 0.20$ 	&$20.83\pm 0.84$	&$12.24$	&$18.54$	\\


MRK~704		&$9.64e+41$	& $17.39\pm 0.22$	&$17.44\pm 0.20$	&$23.79\pm 0.89$	&$13.09$	&$18.64$	\\

		&$2.12e+42$	& $18.10\pm 0.20$	&$17.44\pm 0.20$      	&$20.69\pm 0.18$	&$13.09$	&$18.64$	\\

		&$5.7e+41$	& $16.64\pm 0.20$	&$17.44\pm 0.20$	&$19.87\pm 0.41$	&$13.09$	&$18.64$	\\

AKN~564		&$2.31e+42$	& $17.14\pm 0.23$	&$17.67\pm 0.20$	&$24.70\pm 0.28$	&$12.17$	&$18.92$	\\


NGC~4051	&$4.03e+40$	& $16.86\pm 0.20$	&$16.89\pm 0.20$	&$22.63\pm 0.13$	&$11.77$	&$18.14$	\\

		&$4.80e+40$	& $16.75\pm 0.20$	&$16.89\pm 0.20$	&$18.08\pm 0.28$	&$11.77$	&$18.99$	\\	

NGC~7469	&$2.34e+41$	& $16.82\pm 0.20$	&$17.74\pm 0.20$	&$21.26\pm 0.28$	&$12.57$	&$18.99$	\\	

MRK~766		&$4.36e+41$	& $17.98\pm 0.21$	&$17.09\pm 0.20$	&$21.81\pm 0.21$	&$12.97$	&$18.32$	\\	

		&$1.62e+42$	& $17.61\pm 0.21$	&$17.09\pm 0.20$	&$17.32\pm 0.20$	&$12.97$	&$18.32$	\\	


IRAS~050278	&$1.38e+42$	& $18.08\pm 0.37$	&$17.36\pm 0.20$	&$23.14\pm 0.22$	&$13.33$	&$18.41$	\\	





MR~2251-178	&$7.00e+42$	& $17.62\pm 0.23$	&$17.99\pm 0.20$	&$22.58\pm 0.40$	&$13.97$	&$19.05$	\\	

		&$8.75e+42$	& $17.64\pm 0.20$	&$17.99\pm 0.20$	&$20.52\pm 0.54$	&$13.97$	&$19.05$	\\	


MCG-6-30-15	&$3.66e+41$	& $17.24\pm 0.22$	&$16.82\pm 0.20$	&$22.95\pm 0.16$	&$12.47$	&$18.05$	\\	

		&$2.44e+41$	& $17.81\pm 0.23$	&$16.82\pm 0.20$	&$21.08\pm 0.15$	&$12.47$	&$18.05$	\\	

		&$1.37e+41$	& $16.37\pm 0.20$	&$16.82\pm 0.20$	&$19.45\pm 0.06$	&$12.47$	&$18.05$	\\	

MRK~509		&$9.48e+41$	& $16.84\pm 0.20$	&$17.97\pm 0.20$	&$21.34\pm 0.16$	&$13.77$	&$19.16$	\\	

NGC~3516	&$1.00e+41$	& $17.11\pm 0.21$	&$17.44\pm 0.20$	&$19.41\pm 0.14$	&$13.17$	&$18.21$	\\	

NGC~5548	&$1.05e+42$	& $17.11\pm 0.20$	&$17.48\pm 0.20$	&$19.89\pm 0.12$	&$13.57$	&$18.63$	\\

		&$2.25e+42$	& $18.04\pm 0.20$	&$17.48\pm 0.20$	&$21.33\pm 0.10$	&$13.57$	&$18.63$	\\

NGC~3783	&$2.38e+42$	& $17.49\pm 0.20$	&$17.53\pm 0.20$	&$21.29\pm 0.10$	&$13.27$	&$18.74$	\\

		&$1.14e+42$	& $17.49\pm 0.20$	&$17.53\pm 0.20$	&$19.44\pm 0.10$	&$13.27$	&$18.74$	\\

IC~4329A	&$1.78e+42$	& $18.31\pm 0.21$	&$17.85\pm 0.20$	&$24.89\pm 0.28$	&$13.67$	&$19.11$	\\

		&$5.94e+41$	& $18.68\pm 0.22$	&$17.85\pm 0.20$	&$22.86\pm 0.20$	&$13.67$	&$19.11$	\\

		&$5.94e+41$	& $18.33\pm 0.21$	&$17.85\pm 0.20$	&$20.67\pm 0.21$	&$13.67$	&$19.11$	\\

\hline \hline
\end{tabular} \\ 

{$^1$ The maximum extent of a WA, $\rmax$, is calculated assuming $\vf=1$. \\}

}

\end{table*}


\begin{table*}

{\footnotesize
\centering
\caption{The minimum, dust and maximum mass outflow rates of the warm absorber components of the WAX sample of sources.} \label{Table:mout} 
  \begin{tabular}{lllllllllllllll} \hline\hline 

Sources & $\log\moutmin$ 	&$\log\moutdust$	&$\log\moutmax^1$		&$\log\medd$		\\ 
	& ($\rm gm \,s^{-1}$)	& ($\rm gm\, s^{-1}$)	&($\rm gm \,s^{-1}$) 	& ($\rm gm \,s^{-1}$)	&\\ \hline \\ 

NGC~4593	&$22.91\pm 0.36$	&$22.76\pm 0.31$ 	&$26.27\pm 0.87$	&$24.93$		      	\\


MRK~704	 	& $22.66\pm 0.83$	&$22.72\pm 0.82$	&$29.06\pm 1.20$	&$25.78$	\\

		& $24.06\pm 0.22$	&$23.40\pm 0.22$      	&$26.64\pm 0.20$	&$25.78$\\

		& $23.32\pm 0.22$	&$24.12\pm 0.22$	&$26.55\pm 0.42$	&$25.78$\\

AKN~564		& $22.09\pm 0.32$	&$22.61\pm 0.29$	&$29.65\pm 0.35$	&$24.86$\\


NGC~4051	& $21.87\pm 0.20$	&$21.89\pm 0.21$	&$27.64\pm 0.13$	&$24.46$\\

		& $23.77\pm 0.28$	&$23.91\pm 0.28$	&$25.11\pm 0.34$	&$24.46$\\	

NGC~7469	& $22.78\pm 0.28$	&$23.70\pm 0.28$	&$27.22\pm 0.34$	&$25.26$\\	

MRK~766		& $23.13\pm 0.22$	&$22.24\pm 0.21$	&$26.96\pm 0.22$	&$25.66$\\	

		& $22.87\pm 0.22$	&$22.36\pm 0.21$	&$29.44\pm 0.18$	&$25.66$\\	


IRAS~050278	& $24.03\pm 0.44$	&$23.31\pm 0.31$	&$29.09\pm 0.33$	&$26.02$\\	





MR~2251-178	& $23.88\pm 0.47$	&$24.25\pm 0.45$	&$28.84\pm 0.57$	&$26.66$	\\	

		& $24.63\pm 0.28$	&$24.98\pm 0.28$	&$27.51\pm 0.57$	&$26.66$\\	


MCG-6-30-15	& $22.53\pm 0.27$	&$22.11\pm 0.25$	&$28.24\pm 0.22$	&$25.16$\\	

		& $23.09\pm 0.24$	&$22.09\pm 0.21$	&$26.35\pm 0.17$	&$25.16$\\	

		& $22.72\pm 0.20$	&$23.17\pm 0.20$	&$25.79\pm 0.07$	&$25.16$\\	

MRK~509		& $23.23\pm 0.25$	&$24.36\pm 0.25$	&$27.73\pm 0.21$	&$26.46$\\	

NGC~3516	& $23.63\pm 0.23$	&$23.97\pm 0.22$	&$25.94\pm 0.18$	&$25.86$\\	

NGC~5548	& $23.99\pm 0.22$	&$24.37\pm 0.22$	&$26.77\pm 0.16$	&$26.26$\\

		& $24.06\pm 0.21$	&$23.50\pm 0.21$	&$27.35\pm 0.13$	&$26.26$\\

NGC~3783	& $24.19\pm 0.20$	&$24.23\pm 0.20$	&$28.00\pm 0.11$	&$25.96$\\

		& $24.69\pm 0.23$	&$24.72\pm 0.23$	&$26.64\pm 0.15$	&$25.96$\\

IC~4329A	& $24.07\pm 0.29$	&$23.61\pm 0.29$	&$30.66\pm 0.35$	&$26.36$\\

		& $23.84\pm 0.24$	&$23.01\pm 0.22$	&$28.02\pm 0.22$	&$26.36$\\

		& $24.39\pm 0.29$	&$23.92\pm 0.28$	&$26.74\pm 0.29$	&$26.36$\\

\hline \hline
\end{tabular} \\ 

{$^1$ calculated assuming $\vf=1$. \\}

}

\end{table*}


\begin{table*}

{\footnotesize
\centering
\caption{The minimum, dust and maximum momentum outflow rates of the warm absorber components of the WAX sample of sources.} \label{Table:pout} 
  \begin{tabular}{lllllllllllllll} \hline\hline 

Sources & $\log\poutmin$ 	&$\log\poutdust$	&$\log\poutmax^1$		&$\log\prad$		\\ 
	& ($\rm gm\, cm\, s^{-1}$)	& ($\rm gm\, cm\,s^{-1}$)	&($\rm gm\, cm \,s^{-1}$) 	& ($\rm gm\, cm \, s^{-1}$)	&\\ \hline \\ 

NGC~4593	&$30.62\pm 0.38$	&$30.47\pm 0.34$ 	&$33.98\pm 0.88$	&$33.66$		      	\\


MRK~704	 	& $30.84\pm 0.83$	&$30.89\pm 0.82$	&$37.24\pm 1.20$	&$33.86$	\\

		& $31.88\pm 0.22$	&$31.22\pm 0.22$      	&$34.46\pm 0.20$	&$33.86$\\

		& $31.87\pm 0.23$	&$32.67\pm 0.22$	&$35.10\pm 0.42$	&$33.86$\\

AKN~564		& $29.93\pm 0.33$	&$30.45\pm 0.30$	&$37.49\pm 0.35$	&$34.42$\\


NGC~4051	& $29.65\pm 0.21$	&$29.67\pm 0.21$	&$35.41\pm 0.13$	&$32.86$\\

		& $31.61\pm 0.28$	&$31.75\pm 0.28$	&$32.94\pm 0.34$	&$32.86$\\	

NGC~7469	& $30.98\pm 0.28$	&$31.90\pm 0.28$	&$35.42\pm 0.34$	&$34.55$\\	

MRK~766		& $30.95\pm 0.22$	&$30.06\pm 0.21$	&$34.78\pm 0.22$	&$33.22$\\	

		& $30.88\pm 0.22$	&$30.36\pm 0.21$	&$37.45\pm 0.18$	&$33.22$\\	


IRAS~050278	& $31.98\pm 0.49$	&$31.27\pm 0.31$	&$37.04\pm 0.39$	&$33.39$\\	





MR~2251-178	& $32.38\pm 0.47$	&$32.75\pm 0.45$	&$37.34\pm 0.57$	&$34.67$	\\	

		& $33.12\pm 0.28$	&$33.47\pm 0.28$	&$36.00\pm 0.57$	&$34.67$\\	


MCG-6-30-15	& $30.47\pm 0.28$	&$30.05\pm 0.25$	&$36.18\pm 0.23$	&$32.67$\\	

		& $30.74\pm 0.26$	&$29.75\pm 0.21$	&$34.00\pm 0.19$	&$32.67$\\	

		& $31.09\pm 0.20$	&$31.54\pm 0.20$	&$34.16\pm 0.07$	&$32.67$\\	

MRK~509		& $32.02\pm 0.25$	&$33.15\pm 0.25$	&$36.52\pm 0.21$	&$34.90$\\	

NGC~3516	& $31.99\pm 0.24$	&$32.32\pm 0.23$	&$34.29\pm 0.18$	&$32.99$\\	

NGC~5548	& $32.55\pm 0.22$	&$32.93\pm 0.23$	&$35.33\pm 0.16$	&$33.84$\\

		& $32.15\pm 0.21$	&$31.60\pm 0.21$	&$35.44\pm 0.13$	&$33.84$\\

NGC~3783	& $32.41\pm 0.20$	&$32.45\pm 0.20$	&$36.22\pm 0.11$	&$34.06$\\

		& $32.90\pm 0.23$	&$32.93\pm 0.23$	&$34.85\pm 0.15$	&$34.06$\\

IC~4329A	& $32.08\pm 0.29$	&$31.62\pm 0.29$	&$38.67\pm 0.35$	&$34.80$\\

		& $31.66\pm 0.25$	&$30.83\pm 0.22$	&$35.84\pm 0.22$	&$34.80$\\

		& $32.39\pm 0.29$	&$31.91\pm 0.28$	&$34.73\pm 0.29$	&$34.80$\\

\hline \hline
\end{tabular} \\ 

{$^1$ calculated assuming $\vf=1$. \\}
}

\end{table*}


\begin{table*}

{\footnotesize
\centering
\caption{The minimum, dust and maximum kinetic energy outflow rates of the warm absorber components of the WAX sample of sources.} \label{Table:Ek} 
  \begin{tabular}{lllllllllllllll} \hline\hline 

Sources & $\log\Ekmin$ 	&$\log\Ekdust$	&$\log\Ekmax^1$		&$V_f$		\\ 
	& ($\lunit$)	& ($\lunit$)	&($\lunit$) 		& 	&\\ \hline \\ 

NGC~4593	&$38.03\pm 0.40$	&$37.87\pm 0.36$ 	&$41.38\pm 0.89$	&$0.013$		      	\\


MRK~704 	& $38.72\pm 0.83$	&$38.77\pm 0.83$	&$45.11\pm 1.20$	&$\rm 7.36E-06$	\\

		& $39.40\pm 0.22$	&$38.74\pm 0.22$      	&$41.98\pm 0.21$	&$0.018$\\

		& $40.12\pm 0.23$	&$40.92\pm 0.22$	&$43.35\pm 0.42$	&$0.003$\\

AKN~564		& $37.47\pm 0.34$	&$37.99\pm 0.32$	&$45.03\pm 0.37$	&$\rm 1.10E-05$\\


NGC~4051	& $37.13\pm 0.21$	&$37.15\pm 0.20$	&$42.89\pm 0.13$	&$\rm 3.10E-05$\\

		& $39.15\pm 0.28$	&$39.29\pm 0.28$	&$40.48\pm 0.34$	&$0.385$\\	

NGC~7469	& $38.88\pm 0.29$	&$39.81\pm 0.28$	&$43.32\pm 0.34$	&$0.002$\\	

MRK~766		& $38.47\pm 0.22$	&$37.57\pm 0.22$	&$42.30\pm0.22$		&$0.0009$\\	

		& $38.59\pm 0.22$	&$38.07\pm 0.22$	&$45.16\pm 0.19$	&$\rm 6.22E-06$\\	


IRAS~050278	& $39.63\pm 0.54$	&$38.92\pm 0.44$	&$44.70\pm 0.45$	&$\rm 1.97E-05$\\	





MR~2251-178	& $40.58\pm 0.47$	&$40.95\pm 0.47$	&$45.54\pm 0.58$	&$\rm 7.17E-05$	\\	

		& $41.31\pm 0.28$	&$41.66\pm 0.29$	&$44.19\pm 0.57$	&$0.005$\\	


MCG-6-30-15	& $38.11\pm 0.28$	&$37.69\pm 0.28$	&$43.82\pm 0.23$	&$\rm 2.65E-05$\\	

		& $38.09\pm 0.26$	&$37.10\pm 0.24$	&$41.35\pm 0.19$	&$0.003$\\	

		& $39.16\pm 0.20$	&$39.61\pm 0.20$	&$42.24\pm 0.07$	&$0.002$\\	

MRK~509		& $40.51\pm 0.25$	&$41.64\pm 0.25$	&$45.01\pm 0.21$	&$0.0003$\\	

NGC~3516	& $40.05\pm 0.24$	&$40.38\pm 0.23$	&$42.35\pm 0.18$	&$0.0028$\\	

NGC~5548	& $40.81\pm 0.23$	&$41.18\pm 0.23$	&$43.59\pm 0.16$	&$0.0026$\\

		& $39.94\pm 0.21$	&$39.39\pm 0.21$	&$43.23\pm 0.13$	&$0.0015$\\

NGC~3783	& $40.33\pm 0.20$	&$40.37\pm 0.20$	&$44.14\pm 0.11$	&$0.0008$\\

		& $40.81\pm 0.23$	&$40.85\pm 0.23$	&$42.76\pm 0.15$	&$0.049$\\

IC~4329A	& $39.79\pm 0.30$	&$39.33\pm 0.29$	&$46.38\pm 0.35$	&$\rm 2.85E-06$\\

		& $39.18\pm 0.25$	&$38.34\pm 0.22$	&$43.36\pm 0.22$	&$0.0007$\\

		& $40.08\pm 0.30$	&$39.60\pm 0.30$	&$42.43\pm 0.29$	&$0.044$\\

\hline \hline
\end{tabular} \\ 

{$^1$ calculated assuming $\vf=1$. \\}
}

\end{table*}

\begin{table*}

{\footnotesize
\centering
\caption{The WA distances calculated following the WA `variability' in response to continuum changes. } \label{Table:WA-var} 
  \begin{tabular}{lllllllllllllll} \hline\hline 

Sources	& References			& $\log\xi$		&$\log\nh$				&$n_{\rm e}$	& WA-distance 				& \\ 

	&				& $(\xiunit)$		&$(\cmsqi)$				&$(\cmcubei)$	&$(\cm)$				 \\ \hline \\

NGC~4051&\citet{2012ApJ...746....2K}	&$4.5\pm 0.06$		&$8.10_{-0.99}^{+0.92}\times 10^{20}$	&$10^{11}$	&$1.78_{-0.12}^{+0.12}\times 10^{13}$	&	 \\ 

	&				&$3.28\pm0.05$		&$10.10_{-0.95}^{+0.63}\times10^{20}$	&$10^{11}$	&$7.25_{-0.38}^{+0.29}\times 10^{13}$ \\

	&				&$1.00_{-0.15}^{+0.06}$	&$1.97_{-0.25}^{+0.27}\times10^{20}$	&$10^{10}$	&$317_{-20}^{+58}\times10^{13}$	\\ \\

	&\citet{2007ApJ...659.1022K}$^1$&$-0.26_{-0.08}^{+0.08}$&$10^{21.42}$ 				&$0.58-2.1\times10^{7}$	&$1.3-2.6\times 10^{15}$\\

	&				&$-1.41_{-0.11}^{+0.10}$&$10^{20.73}$				&$>8.1\times10^{7}$	&$<8.9\times 10^{15}$\\ \\

MRK~509& Kaastra et. al. 2012		&$2.01_{-0.02}^{+0.02}$	&$4.8_{-0.4}^{+0.4}\times 10^{24}$	&$<0.28\times 10^{15}$	& $>71\pc$ & \\
	&
					&$2.79_{-0.06}^{+0.06}$	&$5.7_{-0.9}^{+0.9}\times 10^{24}$		&$<10.6\times 10^{15}$	& $>4.7\pc$ & \\
	&
					&$3.60_{-0.27}^{+0.27}$	&$54_{-73}^{+73}\times 10^{24}$		&$<1.7\times 10^{15}$	& $>4.6\pc$ & \\ \\

NGC~3516&\citet{2014ApJ...793...61H}		&$0.43-0.96_{-0.17}^{+0.17}$	&$0.13-0.34_{-0.01}^{+0.01}\times 10^{22}$	&$7.8\times 10^{5}-1.9\times10^{6}$	& $(1.8-2.9)\times 10^{18}$ & \\

	&					&$1.66-2.15_{-0.06}^{+0.06}$	&$0.25-0.40_{-0.01}^{+0.01}\times 10^{22}$	&$>4.9\times 10^{6}$	& $<0.11\pc$ & \\

	&					&$3.26-3.95_{-0.01}^{+0.01}$	&$1.66-3.31_{-0.01}^{+0.01}\times 10^{22}$	&$>6.4\times 10^{4}$	& $\sim 0.13\pc$ & \\ \\

NGC~3783&\citet{2003ApJ...599..933N}		&$-0.9_{-0.10}^{+0.10}$		&$10^{21.9}$					&$<5\times10^{4}$	& $>3.2\pc$ & \\

	&					&$0.3_{-0.01}^{+0.01}$		&$10^{22}$					&$<2.5\times 10^{5}$	& $>0.63\pc$ & \\

	&					&$0.9_{-0.01}^{+0.01}$		&$10^{22.3}$					&$<2.5\times 10^{5}$	& $> 0.18\pc$ & \\ \\

MR~2251-178&\citet{2013ApJ...776...99R}		&$1.27_{-0.02}^{+0.02}$		&$2.12_{-0.07}^{+0.07}\times10^{21}$		&$\sim 3.8\times10^{4}$	& $9-17\pc$ & \\ \\

\hline \hline
\end{tabular} \\ 

{$^1$ The references where the definition of the ionisation parameter is quoted in terms of $U$, we have converted it to $\xi$ by a rough relation $\log\xi=\log U+1.5$ \citep{2012ApJ...753...75C} \\}
{The authors have used two different methods to estimate the WA density and hence the WA distance from the ionising source. For birght sources like NGC~3783 individual absorption line strengths could be measured. Thus, the ratio of the two `density sensitive' absorption lines from the same ionic species gave an estimate of the density. For other sources WA density was measured assuming that the WA responds to a change in the incident continuum variation at a time scale which is of the order of its recombination time scale. The recombination time scale inversely scales with respect to the electron density { \citet{1999ApJ...512..184N}.}}

}

\end{table*}

\clearpage
\appendix

\section{Comparing the mass outflow rates}\label{appendix}

{ In this section we compare the mass outflow rate estimates of the WA between three different methods commonly used in published literature. In our analysis we have used the mass outflow rate estimate from \citet{2007ApJ...659.1022K}, see Equation \ref{equ:mout}. An estimate of the mass outflow rate can be obtained following \citet{2012ApJ...753...75C}, given by 
\begin{equation}\label{crenshaw}
\mout=4 \, \pi \, r \, \nh \, \mu \, m_{\rm p} \, C_{\rm g} \, \vout,
\end{equation}

\noindent where $C_{\rm g}$ is the global covering factor, typically assumed to be $\sim 0.5$. Another estimate of mass outflow rate can be obtained following \citet{2005A&A...431..111B} where the authors assume an inverse square mass density profile of the warm absorber clouds ($n(r) \propto 1/r^2$), and the mass outflow rate in such a scenario is given by,
\begin{equation}\label{blustin}
\mout= \frac{1.23 \, m_{\rm p} \, \lion \, C_v \, \vout \, \Omega}{\xi},
\end{equation}

\noindent where $C_v$ is the volume filling factor and $\Omega$ is the covering fraction. Note that this estimate of mass outflow rate does not explicitly depend on the distance of the absorber $r$, by the special choice of the density profile. 

We calculated the `maximum' mass outflow rates using equations \ref{equ:mout}, \ref{crenshaw} and \ref{blustin}, and using the maximum distance estimate $\rmax$, with a volume filling factor $\vf=1$. The values obtained in the three cases are $\log\mout=30.66\pm 0.35$, $\log\mout=30.87\pm 0.33$ and $\log\mout=29.88 \pm 0.31$ respectively. We find that the mass outflow rate estimates following \citet{2007ApJ...659.1022K} and \citet{2012ApJ...753...75C} are same within errors, while the estimate following \citet{2005A&A...431..111B} is an order of magnitude lower. We should remember that the density profile assumed by \citet{2005A&A...431..111B} is not the true picture for the warm absorber clouds, as we found in Section \ref{subsec:amd} the mass density profile of the WA clouds is $n\propto r^{-1.236}$, and therefore we need to consider this value with caution. From this analysis, we can conclude that the mass outflow rate estimated in this paper following \citet{2007ApJ...659.1022K} is comparable to other geometrical estimates upto an order of magnitude.}


\clearpage

\bibliographystyle{mn2e} 
\bibliography{mybib.bib}

\begin{thebibliography}{}

\bibitem[\protect\citeauthoryear{{Akritas} \& {Bershady}}{{Akritas} \&
  {Bershady}}{1996}]{1996ApJ...470..706A}
{Akritas} M.~G.,  {Bershady} M.~A.,  1996, \apj, 470, 706

\bibitem[\protect\citeauthoryear{{Balogh}, {Pearce}, {Bower} \& {Kay}}{{Balogh}
  et~al.}{2001}]{2001MNRAS.326.1228B}
{Balogh} M.~L.,  {Pearce} F.~R.,  {Bower} R.~G.,    {Kay} S.~T.,  2001, \mnras,
  326, 1228

\bibitem[\protect\citeauthoryear{{Barvainis}}{{Barvainis}}{1987}]{1987ApJ...320..537B}
{Barvainis} R.,  1987, \apj, 320, 537

\bibitem[\protect\citeauthoryear{{Behar}}{{Behar}}{2009}]{2009ApJ...703.1346B}
{Behar} E.,  2009, \apj, 703, 1346

\bibitem[\protect\citeauthoryear{{Blandford} \& {Payne}}{{Blandford} \&
  {Payne}}{1982}]{1982MNRAS.199..883B}
{Blandford} R.~D.,  {Payne} D.~G.,  1982, \mnras, 199, 883

\bibitem[\protect\citeauthoryear{{Blustin}, {Page}, {Fuerst},
  {Branduardi-Raymont} \& {Ashton}}{{Blustin}
  et~al.}{2005}]{2005A&A...431..111B}
{Blustin} A.~J.,  {Page} M.~J.,  {Fuerst} S.~V.,  {Branduardi-Raymont} G.,
  {Ashton} C.~E.,  2005, \aap, 431, 111

\bibitem[\protect\citeauthoryear{{Bottorff}, {Korista} \&
  {Shlosman}}{{Bottorff} et~al.}{2000}]{2000ApJ...537..134B}
{Bottorff} M.~C.,  {Korista} K.~T.,    {Shlosman} I.,  2000, \apj, 537, 134

\bibitem[\protect\citeauthoryear{{Coffey}, {Longinotti},
  {Rodr{\'{\i}}guez-Ardila}, {Guainazzi}, {Miniutti}, {Bianchi}, {de la Calle},
  {Piconcelli}, {Ballo} \& {Linares}}{{Coffey}
  et~al.}{2014}]{2014MNRAS.443.1788C}
{Coffey} D.,  {Longinotti} A.~L.,  {Rodr{\'{\i}}guez-Ardila} A.,  {Guainazzi}
  M.,  {Miniutti} G.,  {Bianchi} S.,  {de la Calle} I.,  {Piconcelli} E.,
  {Ballo} L.,    {Linares} M.,  2014, \mnras, 443, 1788

\bibitem[\protect\citeauthoryear{{Crenshaw} \& {Kraemer}}{{Crenshaw} \&
  {Kraemer}}{2012}]{2012ApJ...753...75C}
{Crenshaw} D.~M.,  {Kraemer} S.~B.,  2012, \apj, 753, 75

\bibitem[\protect\citeauthoryear{{Crenshaw}, {Kraemer}, {Gabel}, {Kaastra},
  {Steenbrugge}, {Brinkman}, {Dunn}, {George}, {Liedahl}, {Paerels}, {Turner}
  \& {Yaqoob}}{{Crenshaw} et~al.}{2003}]{2003ApJ...594..116C}
{Crenshaw} D.~M.,  {Kraemer} S.~B.,  {Gabel} J.~R.,  {Kaastra} J.~S.,
  {Steenbrugge} K.~C.,  {Brinkman} A.~C.,  {Dunn} J.~P.,  {George} I.~M.,
  {Liedahl} D.~A.,  {Paerels} F.~B.~S.,  {Turner} T.~J.,    {Yaqoob} T.,  2003,
  \apj, 594, 116

\bibitem[\protect\citeauthoryear{{Czerny} \& {Hryniewicz}}{{Czerny} \&
  {Hryniewicz}}{2011}]{2011A&A...525L...8C}
{Czerny} B.,  {Hryniewicz} K.,  2011, \aap, 525, L8

\bibitem[\protect\citeauthoryear{{Di Matteo}, {Springel} \& {Hernquist}}{{Di
  Matteo} et~al.}{2005}]{2005Natur.433..604D}
{Di Matteo} T.,  {Springel} V.,    {Hernquist} L.,  2005, \nat, 433, 604

\bibitem[\protect\citeauthoryear{{Elvis}}{{Elvis}}{2000}]{2000ApJ...545...63E}
{Elvis} M.,  2000, \apj, 545, 63

\bibitem[\protect\citeauthoryear{{Everett} \& {Ballantyne}}{{Everett} \&
  {Ballantyne}}{2004}]{2004ApJ...615L..13E}
{Everett} J.~E.,  {Ballantyne} D.~R.,  2004, \apjl, 615, L13

\bibitem[\protect\citeauthoryear{{Fabian}, {Vasudevan} \& {Gandhi}}{{Fabian}
  et~al.}{2008}]{2008MNRAS.385L..43F}
{Fabian} A.~C.,  {Vasudevan} R.~V.,    {Gandhi} P.,  2008, \mnras, 385, L43

\bibitem[\protect\citeauthoryear{{Fukumura}, {Kazanas}, {Contopoulos} \&
  {Behar}}{{Fukumura} et~al.}{2010a}]{2010ApJ...715..636F}
{Fukumura} K.,  {Kazanas} D.,  {Contopoulos} I.,    {Behar} E.,  2010a, \apj,
  715, 636

\bibitem[\protect\citeauthoryear{{Fukumura}, {Kazanas}, {Contopoulos} \&
  {Behar}}{{Fukumura} et~al.}{2010b}]{2010ApJ...723L.228F}
{Fukumura} K.,  {Kazanas} D.,  {Contopoulos} I.,    {Behar} E.,  2010b, \apjl,
  723, L228

\bibitem[\protect\citeauthoryear{{Gallagher}, {Everett}, {Abado} \&
  {Keating}}{{Gallagher} et~al.}{2015}]{2015MNRAS.451.2991G}
{Gallagher} S.~C.,  {Everett} J.~E.,  {Abado} M.~M.,    {Keating} S.~K.,  2015,
  \mnras, 451, 2991

\bibitem[\protect\citeauthoryear{{Gaspari}, {Brighenti} \& {Temi}}{{Gaspari}
  et~al.}{2012}]{2012MNRAS.424..190G}
{Gaspari} M.,  {Brighenti} F.,    {Temi} P.,  2012, \mnras, 424, 190

\bibitem[\protect\citeauthoryear{{Gaspari}, {Melioli}, {Brighenti} \&
  {D'Ercole}}{{Gaspari} et~al.}{2011}]{2011MNRAS.411..349G}
{Gaspari} M.,  {Melioli} C.,  {Brighenti} F.,    {D'Ercole} A.,  2011, \mnras,
  411, 349

\bibitem[\protect\citeauthoryear{{Giustini} \& {Proga}}{{Giustini} \&
  {Proga}}{2012}]{2012ApJ...758...70G}
{Giustini} M.,  {Proga} D.,  2012, \apj, 758, 70

\bibitem[\protect\citeauthoryear{{Gofford}, {Reeves}, {McLaughlin}, {Braito},
  {Turner}, {Tombesi} \& {Cappi}}{{Gofford}
  et~al.}{2015a}]{2015arXiv150600614G}
{Gofford} J.,  {Reeves} J.~N.,  {McLaughlin} D.~E.,  {Braito} V.,  {Turner}
  T.~J.,  {Tombesi} F.,    {Cappi} M.,  2015a, ArXiv e-prints

\bibitem[\protect\citeauthoryear{{Gofford}, {Reeves}, {McLaughlin}, {Braito},
  {Turner}, {Tombesi} \& {Cappi}}{{Gofford}
  et~al.}{2015b}]{2015MNRAS.451.4169G}
{Gofford} J.,  {Reeves} J.~N.,  {McLaughlin} D.~E.,  {Braito} V.,  {Turner}
  T.~J.,  {Tombesi} F.,    {Cappi} M.,  2015b, \mnras, 451, 4169

\bibitem[\protect\citeauthoryear{{Gofford}, {Reeves}, {Turner}, {Tombesi},
  {Braito}, {Porquet}, {Miller}, {Kraemer} \& {Fukazawa}}{{Gofford}
  et~al.}{2011}]{2011MNRAS.414.3307G}
{Gofford} J.,  {Reeves} J.~N.,  {Turner} T.~J.,  {Tombesi} F.,  {Braito} V.,
  {Porquet} D.,  {Miller} L.,  {Kraemer} S.~B.,    {Fukazawa} Y.,  2011,
  \mnras, 414, 3307

\bibitem[\protect\citeauthoryear{{Higginbottom}, {Proga}, {Knigge}, {Long},
  {Matthews} \& {Sim}}{{Higginbottom} et~al.}{2014}]{2014ApJ...789...19H}
{Higginbottom} N.,  {Proga} D.,  {Knigge} C.,  {Long} K.~S.,  {Matthews} J.~H.,
     {Sim} S.~A.,  2014, \apj, 789, 19

\bibitem[\protect\citeauthoryear{{Holczer}, {Behar} \& {Kaspi}}{{Holczer}
  et~al.}{2005}]{2005ApJ...632..788H}
{Holczer} T.,  {Behar} E.,    {Kaspi} S.,  2005, \apj, 632, 788

\bibitem[\protect\citeauthoryear{{Holczer}, {Behar} \& {Kaspi}}{{Holczer}
  et~al.}{2007}]{2007ApJ...663..799H}
{Holczer} T.,  {Behar} E.,    {Kaspi} S.,  2007, \apj, 663, 799

\bibitem[\protect\citeauthoryear{{H{\"o}nig} \& {Beckert}}{{H{\"o}nig} \&
  {Beckert}}{2007}]{2007MNRAS.380.1172H}
{H{\"o}nig} S.~F.,  {Beckert} T.,  2007, \mnras, 380, 1172

\bibitem[\protect\citeauthoryear{{Hopkins} \& {Elvis}}{{Hopkins} \&
  {Elvis}}{2010}]{2010MNRAS.401....7H}
{Hopkins} P.~F.,  {Elvis} M.,  2010, \mnras, 401, 7

\bibitem[\protect\citeauthoryear{{Huerta}, {Krongold}, {Nicastro}, {Mathur},
  {Longinotti} \& {Jimenez-Bailon}}{{Huerta}
  et~al.}{2014}]{2014ApJ...793...61H}
{Huerta} E.~M.,  {Krongold} Y.,  {Nicastro} F.,  {Mathur} S.,  {Longinotti}
  A.~L.,    {Jimenez-Bailon} E.,  2014, \apj, 793, 61

\bibitem[\protect\citeauthoryear{{Ishibashi} \& {Fabian}}{{Ishibashi} \&
  {Fabian}}{2015}]{2015MNRAS.451...93I}
{Ishibashi} W.,  {Fabian} A.~C.,  2015, \mnras, 451, 93

\bibitem[\protect\citeauthoryear{{Kaastra}, {Detmers}, {Mehdipour}, {Arav},
  {Behar}, {Bianchi}, {Branduardi-Raymont}, {Cappi}, {Costantini}, {Ebrero},
  {Kriss}, {Paltani}, {Petrucci}, {Pinto}, {Ponti}, {Steenbrugge} \& {de
  Vries}}{{Kaastra} et~al.}{2012}]{2012A&A...539A.117K}
{Kaastra} J.~S.,  {Detmers} R.~G.,  {Mehdipour} M.,  {Arav} N.,  {Behar} E.,
  {Bianchi} S.,  {Branduardi-Raymont} G.,  {Cappi} M.,  {Costantini} E.,
  {Ebrero} J.,  {Kriss} G.~A.,  {Paltani} S.,  {Petrucci} P.-O.,  {Pinto} C.,
  {Ponti} G.,  {Steenbrugge} K.~C.,    {de Vries} C.~P.,  2012, \aap, 539, A117

\bibitem[\protect\citeauthoryear{{Kaastra}, {Steenbrugge}, {Raassen}, {van der
  Meer}, {Brinkman}, {Liedahl}, {Behar} \& {de Rosa}}{{Kaastra}
  et~al.}{2002}]{2002A&A...386..427K}
{Kaastra} J.~S.,  {Steenbrugge} K.~C.,  {Raassen} A.~J.~J.,  {van der Meer}
  R.~L.~J.,  {Brinkman} A.~C.,  {Liedahl} D.~A.,  {Behar} E.,    {de Rosa} A.,
  2002, \aap, 386, 427

\bibitem[\protect\citeauthoryear{{Kaspi}, {Netzer}, {Chelouche}, {George},
  {Nandra} \& {Turner}}{{Kaspi} et~al.}{2004}]{2004ApJ...611...68K}
{Kaspi} S.,  {Netzer} H.,  {Chelouche} D.,  {George} I.~M.,  {Nandra} K.,
  {Turner} T.~J.,  2004, \apj, 611, 68

\bibitem[\protect\citeauthoryear{{Kazanas}, {Fukumura}, {Behar} \&
  {Contopoulos}}{{Kazanas} et~al.}{2012}]{2012ASPC..460..181K}
{Kazanas} D.,  {Fukumura} K.,  {Behar} E.,    {Contopoulos} I.,  2012, in
  {Chartas} G.,  {Hamann} F.,   {Leighly} K.~M.,  eds, AGN Winds in Charleston
  Vol.~460 of Astronomical Society of the Pacific Conference Series, {X-ray
  Absorbers, MHD Winds, and AGN Unification}.
p.~181

\bibitem[\protect\citeauthoryear{{King}, {Miller} \& {Raymond}}{{King}
  et~al.}{2012}]{2012ApJ...746....2K}
{King} A.~L.,  {Miller} J.~M.,    {Raymond} J.,  2012, \apj, 746, 2

\bibitem[\protect\citeauthoryear{{King}, {Miller}, {Raymond}, {Fabian},
  {Reynolds}, {G{\"u}ltekin}, {Cackett}, {Allen}, {Proga} \& {Kallman}}{{King}
  et~al.}{2013}]{2013ApJ...762..103K}
{King} A.~L.,  {Miller} J.~M.,  {Raymond} J.,  {Fabian} A.~C.,  {Reynolds}
  C.~S.,  {G{\"u}ltekin} K.,  {Cackett} E.~M.,  {Allen} S.~W.,  {Proga} D.,
  {Kallman} T.~R.,  2013, \apj, 762, 103

\bibitem[\protect\citeauthoryear{{King}, {Miller}, {Raymond}, {Fabian},
  {Reynolds}, {Kallman}, {Maitra}, {Cackett} \& {Rupen}}{{King}
  et~al.}{2012}]{2012ApJ...746L..20K}
{King} A.~L.,  {Miller} J.~M.,  {Raymond} J.,  {Fabian} A.~C.,  {Reynolds}
  C.~S.,  {Kallman} T.~R.,  {Maitra} D.,  {Cackett} E.~M.,    {Rupen} M.~P.,
  2012, \apjl, 746, L20

\bibitem[\protect\citeauthoryear{{King}}{{King}}{2010}]{2010MNRAS.402.1516K}
{King} A.~R.,  2010, \mnras, 402, 1516

\bibitem[\protect\citeauthoryear{{Kinkhabwala}, {Sako}, {Behar}, {Kahn},
  {Paerels}, {Brinkman}, {Kaastra}, {Gu} \& {Liedahl}}{{Kinkhabwala}
  et~al.}{2002}]{2002ApJ...575..732K}
{Kinkhabwala} A.,  {Sako} M.,  {Behar} E.,  {Kahn} S.~M.,  {Paerels} F.,
  {Brinkman} A.~C.,  {Kaastra} J.~S.,  {Gu} M.~F.,    {Liedahl} D.~A.,  2002,
  \apj, 575, 732

\bibitem[\protect\citeauthoryear{{Krolik} \& {Kriss}}{{Krolik} \&
  {Kriss}}{2001}]{2001ApJ...561..684K}
{Krolik} J.~H.,  {Kriss} G.~A.,  2001, \apj, 561, 684

\bibitem[\protect\citeauthoryear{{Krongold}, {Nicastro}, {Elvis}, {Brickhouse},
  {Binette}, {Mathur} \& {Jim{\'e}nez-Bail{\'o}n}}{{Krongold}
  et~al.}{2007}]{2007ApJ...659.1022K}
{Krongold} Y.,  {Nicastro} F.,  {Elvis} M.,  {Brickhouse} N.,  {Binette} L.,
  {Mathur} S.,    {Jim{\'e}nez-Bail{\'o}n} E.,  2007, \apj, 659, 1022

\bibitem[\protect\citeauthoryear{{Krongold}, {Nicastro}, {Elvis}, {Brickhouse},
  {Jim{\'e}nez-Bail{\'o}n}, {Binette} \& {Mathur}}{{Krongold}
  et~al.}{2008}]{2008RMxAC..32..123K}
{Krongold} Y.,  {Nicastro} F.,  {Elvis} M.,  {Brickhouse} N.,
  {Jim{\'e}nez-Bail{\'o}n} E.,  {Binette} L.,    {Mathur} S.,  2008, in Revista
  Mexicana de Astronomia y Astrofisica Conference Series Vol.~32 of Revista
  Mexicana de Astronomia y Astrofisica, vol. 27, {Can Ionized Outflows in AGN
  Produce Important Feedback Effects? The Case of NGC 4051}.
pp 123--127

\bibitem[\protect\citeauthoryear{{Krongold}, {Nicastro}, {Elvis}, {Brickhouse},
  {Mathur} \& {Zezas}}{{Krongold} et~al.}{2005}]{2005ApJ...620..165K}
{Krongold} Y.,  {Nicastro} F.,  {Elvis} M.,  {Brickhouse} N.~S.,  {Mathur} S.,
    {Zezas} A.,  2005, \apj, 620, 165

\bibitem[\protect\citeauthoryear{{Laha}, {Dewangan}, {Chakravorty} \&
  {Kembhavi}}{{Laha} et~al.}{2013}]{2013ApJ...777....2L}
{Laha} S.,  {Dewangan} G.~C.,  {Chakravorty} S.,    {Kembhavi} A.~K.,  2013,
  \apj, 777, 2

\bibitem[\protect\citeauthoryear{{Laha}, {Dewangan} \& {Kembhavi}}{{Laha}
  et~al.}{2011}]{2011ApJ...734...75L}
{Laha} S.,  {Dewangan} G.~C.,    {Kembhavi} A.~K.,  2011, \apj, 734, 75

\bibitem[\protect\citeauthoryear{{Laha}, {Guainazzi}, {Dewangan}, {Chakravorty}
  \& {Kembhavi}}{{Laha} et~al.}{2014}]{2014MNRAS.441.2613L}
{Laha} S.,  {Guainazzi} M.,  {Dewangan} G.~C.,  {Chakravorty} S.,    {Kembhavi}
  A.~K.,  2014, \mnras, 441, 2613

\bibitem[\protect\citeauthoryear{{Lee}, {Kriss}, {Chakravorty}, {Rahoui},
  {Young}, {Brandt}, {Hines}, {Ogle} \& {Reynolds}}{{Lee}
  et~al.}{2013}]{2013MNRAS.430.2650L}
{Lee} J.~C.,  {Kriss} G.~A.,  {Chakravorty} S.,  {Rahoui} F.,  {Young} A.~J.,
  {Brandt} W.~N.,  {Hines} D.~C.,  {Ogle} P.~M.,    {Reynolds} C.~S.,  2013,
  \mnras, 430, 2650

\bibitem[\protect\citeauthoryear{{Lee}, {Ogle}, {Canizares}, {Marshall},
  {Schulz}, {Morales}, {Fabian} \& {Iwasawa}}{{Lee}
  et~al.}{2001}]{2001ApJ...554L..13L}
{Lee} J.~C.,  {Ogle} P.~M.,  {Canizares} C.~R.,  {Marshall} H.~L.,  {Schulz}
  N.~S.,  {Morales} R.,  {Fabian} A.~C.,    {Iwasawa} K.,  2001, \apjl, 554,
  L13

\bibitem[\protect\citeauthoryear{{McKernan}, {Yaqoob} \& {Reynolds}}{{McKernan}
  et~al.}{2007}]{2007MNRAS.379.1359M}
{McKernan} B.,  {Yaqoob} T.,    {Reynolds} C.~S.,  2007, \mnras, 379, 1359

\bibitem[\protect\citeauthoryear{{Miller}, {Raymond}, {Fabian}, {Steeghs},
  {Homan}, {Reynolds}, {van der Klis} \& {Wijnands}}{{Miller}
  et~al.}{2006}]{2006Natur.441..953M}
{Miller} J.~M.,  {Raymond} J.,  {Fabian} A.,  {Steeghs} D.,  {Homan} J.,
  {Reynolds} C.,  {van der Klis} M.,    {Wijnands} R.,  2006, \nat, 441, 953

\bibitem[\protect\citeauthoryear{{Miniutti}, {Sanfrutos}, {Beuchert},
  {Ag{\'{\i}}s-Gonz{\'a}lez}, {Longinotti}, {Piconcelli}, {Krongold},
  {Guainazzi}, {Bianchi}, {Matt} \& {Jim{\'e}nez-Bail{\'o}n}}{{Miniutti}
  et~al.}{2014}]{2014MNRAS.437.1776M}
{Miniutti} G.,  {Sanfrutos} M.,  {Beuchert} T.,  {Ag{\'{\i}}s-Gonz{\'a}lez} B.,
   {Longinotti} A.~L.,  {Piconcelli} E.,  {Krongold} Y.,  {Guainazzi} M.,
  {Bianchi} S.,  {Matt} G.,    {Jim{\'e}nez-Bail{\'o}n} E.,  2014, \mnras, 437,
  1776

\bibitem[\protect\citeauthoryear{{Nemmen}, {Georganopoulos}, {Guiriec},
  {Meyer}, {Gehrels} \& {Sambruna}}{{Nemmen}
  et~al.}{2012}]{2012Sci...338.1445N}
{Nemmen} R.~S.,  {Georganopoulos} M.,  {Guiriec} S.,  {Meyer} E.~T.,  {Gehrels}
  N.,    {Sambruna} R.~M.,  2012, Science, 338, 1445

\bibitem[\protect\citeauthoryear{{Netzer}}{{Netzer}}{2015}]{2015ARA&A..53..365N}
{Netzer} H.,  2015, \araa, 53, 365

\bibitem[\protect\citeauthoryear{{Netzer}, {Kaspi}, {Behar}, {Brandt},
  {Chelouche}, {George}, {Crenshaw}, {Gabel}, {Hamann}, {Kraemer}, {Kriss},
  {Nandra}, {Peterson}, {Shields} \& {Turner}}{{Netzer}
  et~al.}{2003}]{2003ApJ...599..933N}
{Netzer} H.,  {Kaspi} S.,  {Behar} E.,  {Brandt} W.~N.,  {Chelouche} D.,
  {George} I.~M.,  {Crenshaw} D.~M.,  {Gabel} J.~R.,  {Hamann} F.~W.,
  {Kraemer} S.~B.,  {Kriss} G.~A.,  {Nandra} K.,  {Peterson} B.~M.,  {Shields}
  J.~C.,    {Turner} T.~J.,  2003, \apj, 599, 933

\bibitem[\protect\citeauthoryear{{Nicastro}, {Fiore}, {Perola} \&
  {Elvis}}{{Nicastro} et~al.}{1999}]{1999ApJ...512..184N}
{Nicastro} F.,  {Fiore} F.,  {Perola} G.~C.,    {Elvis} M.,  1999, \apj, 512,
  184

\bibitem[\protect\citeauthoryear{{Ogle}, {Mason}, {Page}, {Salvi}, {Cordova},
  {McHardy} \& {Priedhorsky}}{{Ogle} et~al.}{2004}]{2004ApJ...606..151O}
{Ogle} P.~M.,  {Mason} K.~O.,  {Page} M.~J.,  {Salvi} N.~J.,  {Cordova} F.~A.,
  {McHardy} I.~M.,    {Priedhorsky} W.~C.,  2004, \apj, 606, 151

\bibitem[\protect\citeauthoryear{{Ponti}, {Miniutti}, {Cappi}, {Maraschi},
  {Fabian} \& {Iwasawa}}{{Ponti} et~al.}{2006}]{2006MNRAS.368..903P}
{Ponti} G.,  {Miniutti} G.,  {Cappi} M.,  {Maraschi} L.,  {Fabian} A.~C.,
  {Iwasawa} K.,  2006, \mnras, 368, 903

\bibitem[\protect\citeauthoryear{{Pounds} \& {King}}{{Pounds} \&
  {King}}{2013}]{2013MNRAS.433.1369P}
{Pounds} K.~A.,  {King} A.~R.,  2013, \mnras, 433, 1369

\bibitem[\protect\citeauthoryear{{Pounds}, {King}, {Page} \&
  {O'Brien}}{{Pounds} et~al.}{2003}]{2003MNRAS.346.1025P}
{Pounds} K.~A.,  {King} A.~R.,  {Page} K.~L.,    {O'Brien} P.~T.,  2003,
  \mnras, 346, 1025

\bibitem[\protect\citeauthoryear{{Proga} \& {Begelman}}{{Proga} \&
  {Begelman}}{2003}]{2003ApJ...582...69P}
{Proga} D.,  {Begelman} M.~C.,  2003, \apj, 582, 69

\bibitem[\protect\citeauthoryear{{Proga} \& {Kallman}}{{Proga} \&
  {Kallman}}{2004}]{2004ApJ...616..688P}
{Proga} D.,  {Kallman} T.~R.,  2004, \apj, 616, 688

\bibitem[\protect\citeauthoryear{{Reeves}, {Porquet}, {Braito}, {Gofford},
  {Nardini}, {Turner}, {Crenshaw} \& {Kraemer}}{{Reeves}
  et~al.}{2013}]{2013ApJ...776...99R}
{Reeves} J.~N.,  {Porquet} D.,  {Braito} V.,  {Gofford} J.,  {Nardini} E.,
  {Turner} T.~J.,  {Crenshaw} D.~M.,    {Kraemer} S.~B.,  2013, \apj, 776, 99

\bibitem[\protect\citeauthoryear{{Revnivtsev}, {Sazonov}, {Jahoda} \&
  {Gilfanov}}{{Revnivtsev} et~al.}{2004}]{2004A&A...418..927R}
{Revnivtsev} M.,  {Sazonov} S.,  {Jahoda} K.,    {Gilfanov} M.,  2004, \aap,
  418, 927

\bibitem[\protect\citeauthoryear{{Sako}, {Kahn}, {Branduardi-Raymont},
  {Kaastra}, {Brinkman}, {Page}, {Behar}, {Paerels}, {Kinkhabwala}, {Liedahl}
  \& {den Herder}}{{Sako} et~al.}{2003}]{2003ApJ...596..114S}
{Sako} M.,  {Kahn} S.~M.,  {Branduardi-Raymont} G.,  {Kaastra} J.~S.,
  {Brinkman} A.~C.,  {Page} M.~J.,  {Behar} E.,  {Paerels} F.,  {Kinkhabwala}
  A.,  {Liedahl} D.~A.,    {den Herder} J.~W.,  2003, \apj, 596, 114

\bibitem[\protect\citeauthoryear{{Sim}, {Proga}, {Miller}, {Long} \&
  {Turner}}{{Sim} et~al.}{2010}]{2010MNRAS.408.1396S}
{Sim} S.~A.,  {Proga} D.,  {Miller} L.,  {Long} K.~S.,    {Turner} T.~J.,
  2010, \mnras, 408, 1396

\bibitem[\protect\citeauthoryear{{Tananbaum}, {Avni}, {Branduardi}, {Elvis},
  {Fabbiano}, {Feigelson}, {Giacconi}, {Henry}, {Pye}, {Soltan} \&
  {Zamorani}}{{Tananbaum} et~al.}{1979}]{1979ApJ...234L...9T}
{Tananbaum} H.,  {Avni} Y.,  {Branduardi} G.,  {Elvis} M.,  {Fabbiano} G.,
  {Feigelson} E.,  {Giacconi} R.,  {Henry} J.~P.,  {Pye} J.~P.,  {Soltan} A.,
   {Zamorani} G.,  1979, \apjl, 234, L9

\bibitem[\protect\citeauthoryear{{Thompson}, {Fabian}, {Quataert} \&
  {Murray}}{{Thompson} et~al.}{2015}]{2015MNRAS.449..147T}
{Thompson} T.~A.,  {Fabian} A.~C.,  {Quataert} E.,    {Murray} N.,  2015,
  \mnras, 449, 147

\bibitem[\protect\citeauthoryear{{Tombesi}, {Cappi}, {Reeves} \&
  {Braito}}{{Tombesi} et~al.}{2012}]{2012MNRAS.422L...1T}
{Tombesi} F.,  {Cappi} M.,  {Reeves} J.~N.,    {Braito} V.,  2012, \mnras, 422,
  L1

\bibitem[\protect\citeauthoryear{{Tombesi}, {Cappi}, {Reeves}, {Nemmen},
  {Braito}, {Gaspari} \& {Reynolds}}{{Tombesi}
  et~al.}{2013}]{2013MNRAS.430.1102T}
{Tombesi} F.,  {Cappi} M.,  {Reeves} J.~N.,  {Nemmen} R.~S.,  {Braito} V.,
  {Gaspari} M.,    {Reynolds} C.~S.,  2013, \mnras, 430, 1102

\bibitem[\protect\citeauthoryear{{Tombesi}, {Cappi}, {Reeves}, {Palumbo},
  {Braito} \& {Dadina}}{{Tombesi} et~al.}{2011}]{2011ApJ...742...44T}
{Tombesi} F.,  {Cappi} M.,  {Reeves} J.~N.,  {Palumbo} G.~G.~C.,  {Braito} V.,
    {Dadina} M.,  2011, \apj, 742, 44

\bibitem[\protect\citeauthoryear{{Tombesi}, {Cappi}, {Reeves}, {Palumbo},
  {Yaqoob}, {Braito} \& {Dadina}}{{Tombesi} et~al.}{2010}]{2010A&A...521A..57T}
{Tombesi} F.,  {Cappi} M.,  {Reeves} J.~N.,  {Palumbo} G.~G.~C.,  {Yaqoob} T.,
  {Braito} V.,    {Dadina} M.,  2010, \aap, 521, A57

\bibitem[\protect\citeauthoryear{{Turner}, {Fabian}, {Lee} \&
  {Vaughan}}{{Turner} et~al.}{2004}]{2004MNRAS.353..319T}
{Turner} A.~K.,  {Fabian} A.~C.,  {Lee} J.~C.,    {Vaughan} S.,  2004, \mnras,
  353, 319

\bibitem[\protect\citeauthoryear{{Urry} \& {Padovani}}{{Urry} \&
  {Padovani}}{1995}]{1995PASP..107..803U}
{Urry} C.~M.,  {Padovani} P.,  1995, \pasp, 107, 803

\bibitem[\protect\citeauthoryear{{Vasudevan}, {Fabian}, {Mushotzky},
  {Mel{\'e}ndez}, {Winter} \& {Trippe}}{{Vasudevan}
  et~al.}{2013}]{2013MNRAS.431.3127V}
{Vasudevan} R.~V.,  {Fabian} A.~C.,  {Mushotzky} R.~F.,  {Mel{\'e}ndez} M.,
  {Winter} L.~M.,    {Trippe} M.~L.,  2013, \mnras, 431, 3127

\bibitem[\protect\citeauthoryear{{Winter}, {Veilleux}, {McKernan} \&
  {Kallman}}{{Winter} et~al.}{2012}]{2012ApJ...745..107W}
{Winter} L.~M.,  {Veilleux} S.,  {McKernan} B.,    {Kallman} T.~R.,  2012,
  \apj, 745, 107

\end{thebibliography}

\end{document}